\documentclass[iop]{emulateapj}
\usepackage{natbib}
\usepackage{graphicx}
\def \lp{\>\> .}
\def \lc{\>\> ,}

\newcommand{\degree}{\mbox{$^{\circ}$}}
\newcommand{\am}{\mbox{\arcmin}}
\newcommand{\as}{\mbox{\arcsec}}

\newcommand{\kms}{\mbox{\,km\,s$^{-1}$}}
\newcommand\cmv{\mbox{cm$^{-3}$}}
\newcommand\cmc{\mbox{cm$^{-2}$}}

\newcommand{\otwo}{O$_2$}

\def\CII{C{\sc ii}}

\def\OI{O{\sc i}}

\newcommand{\peaka}{\mbox{$Peak\,A$}}
\newcommand{\peakone}{\mbox{$H_2\,Peak$ {\it 1}}}
\newcommand{\xotwo}{\mbox{$X$(O$_2$)}}
\newcommand{\Notwo}{\mbox{$N$(O$_2$)}}
\def \vlsr{$V_{\rm LSR}$}        
\def \Ta{$T_{\rm A}$}
\def \Tex{$T_{\rm ex}$}
\def \Eu{$E_{\rm u}$}

\begin{document}

\title {Herschel HIFI observations of O$_2$ toward Orion: special conditions for shock enhanced emission} 
\author{
Jo-Hsin Chen\altaffilmark{1}, 
Paul F. Goldsmith\altaffilmark{1},
Serena Viti\altaffilmark{2},
Ronald Snell\altaffilmark{3},
Dariusz C. Lis\altaffilmark{4},
Arnold Benz\altaffilmark{5},
Edwin Bergin\altaffilmark{6},
John Black\altaffilmark{7},
Paola Caselli\altaffilmark{8},
Pierre Encrenaz\altaffilmark{9},
Edith Falgarone\altaffilmark{9},
Javier R.  Goicoechea\altaffilmark{10},
\AA ke Hjalmarson\altaffilmark{7},
David Hollenbach\altaffilmark{11},
Michael Kaufman\altaffilmark{12},
Gary Melnick\altaffilmark{13},
David Neufeld\altaffilmark{14},
Laurent Pagani\altaffilmark{15},
Floris van der Tak\altaffilmark{16},
Ewine van Dishoeck\altaffilmark{17,8}, and
Umut A. Y{\i}ld{\i}z\altaffilmark{1}
}

\vspace{0.2cm}

\altaffiltext{1}{Jet Propulsion Laboratory, California Institute of Technology, 4800 Oak Grove Drive, Pasadena CA, 91109, USA}
\altaffiltext{2}{Department of Physics and Astronomy, University College London, London WC1E 6BT, United Kingdom}
\altaffiltext{3}{University of Massachusetts, Department of Astronomy, LGRT-B 619E, 710 North Pleasant Street, Amherst, MA 01003, USA}
\altaffiltext{4}{California Institute of Technology, Cahill Center for Astronomy and Astrophysics 301-17, Pasadena, CA 91125, USA and Sorbonne Universit\'{e}s, Universit\'{e} Pierre et Marie Curie, Paris 6, CNRS, Observatoire de Paris, UMR 8112, LERMA, Paris, France}
\altaffiltext{5}{Institute of Astronomy, ETH Zurich, Zurich, Switzerland}
\altaffiltext{6} {Department of Astronomy, University of Michigan, 500 Church Street, Ann Arbor MI 48109}
\altaffiltext{7}  {Department of Earth \& Space Sciences, Chalmers University of Technology, Onsala Space Observatory, SE-439 92 Onsala, Sweden}
\altaffiltext{8} {Max-Planck-Institut f\"{u}r Extraterrestrische Physik, Giessenbachstrasse 1, 85748, Garching, Germany}
\altaffiltext{9}{LRA/LERMA, CNRS, UMR8112, Observatoire de Paris \& \'{E}cole Normale Sup\'{e}rieure, 24 rue Lhomond, 75231 Paris Cedex 05, France}
\altaffiltext{10}{Instituto de Ciencia de Materiales de Madrid (ICMM-CSIC) E-28049, Cantoblanco, Madrid, Spain}
\altaffiltext{11}  {SETI Institute, Mountain View CA 94043}
\altaffiltext{12} {Department of Physics and Astronomy, San Jos\'{e} State University, San Jose CA 95192}
\altaffiltext{13}{Harvard-Smithsonian Center for Astrophysics, 60 Garden Street, MS 66, Cambridge MA 02138}
\altaffiltext{14}{Department of Physics and Astronomy, Johns Hopkins University, 3400 North Charles Street, Baltimore, MD 21218}
\altaffiltext{15}{LERMA \& UMR8112 du CNRS, Observatoire de Paris, 61 Av. de l'Observatoire, 75014, Paris, France}
\altaffiltext{16}{SRON Netherlands Institute for Space Research, PO Box 800, 9700 AV, and Kapteyn Astronomical Institute, University of Groningen, Groningen, The Netherlands}
\altaffiltext{17}{Leiden Observatory, Leiden University, PO Box 9513, 2300 RA, Leiden, The Netherlands}

\clearpage
\newpage

\begin{abstract}

We report observations of molecular oxygen ({\otwo}) rotational transitions at 487 GHz, 774 GHz, and 1121 GHz toward Orion \peaka. The \otwo\ lines at 487 GHz and 774 GHz are detected at velocities of 10--12~\kms\ with line widths $\sim$3~\kms; however, the transition at 1121 GHz is not detected. The observed line characteristics, combined with the results of earlier observations, suggest that the region responsible for the \otwo\ emission is $\simeq$ 9\as\ (6$\times$10$^{16}$ cm) in size, and is located close to the \peakone\ position (where vibrationally--excited H$_2$ emission peaks), and not at \peaka, 23\as\ away.  The peak \otwo\ column density is $\simeq$ 1.1$\times$10$^{18}$ \cmc.  The line velocity is close to that of 621 GHz water maser emission found in this portion of the Orion Molecular Cloud, and having a shock with velocity vector lying nearly in the plane of the sky is consistent with producing maximum maser gain along the line--of--sight.  The enhanced \otwo\ abundance compared to that generally found in dense interstellar clouds can be explained by passage of a low--velocity $C$-shock through a clump with preshock density 2$\times$10$^4$ \cmv, if a reasonable flux of UV radiation is present.  The postshock \otwo\ can explain the emission from  the source if its line of sight dimension is $\simeq$ 10 times larger than its size on the plane of the sky.  The special geometry and conditions required may explain why \otwo\ emission has not been detected in the cores of other massive star--forming molecular clouds. 
\end{abstract}

\section{Introduction}
Oxygen is the third most abundant element in the Universe, but our understanding of its form in the dense interstellar medium is still limited. The molecular form of oxygen, {\otwo}, was thought to be a significant reservoir of this element and an important coolant of the interstellar medium \citep{goldsmith78}. Models incorporating only gas-phase chemistry suggested that the fractional abundance of \otwo, \xotwo = $n$(\otwo)/$n$(H$_2$) = $N$(\otwo)/$N$(H$_2$) for uniform conditions along the line of sight, could be greater than $10^{-5}$ \citep[e.g.][]{bergin95} in well-shielded regions. However, attempts to detect {\otwo} were largely unsuccessful (\citeauthor{goldsmith11} 2011, and references therein). Previous searches carried out with {\it the Submillimeter Wave Astronomy Satellite} ({\it SWAS}; \citeauthor{melnick00} 2000) yielded upper limits of \otwo\ abundance, \xotwo\ $\le$10$^{-7}$, two orders of magnitude below model predictions \citep{goldsmith00}. Observations with the {\it Odin Satellite} \citep{nordh03} gave an upper limit of \xotwo\ $\le10^{-7}$ in cold clouds \citep{pagani03}, with the exception of a single detection toward the $\rho$ Ophiuchi A cloud with \xotwo\ $\sim$5$\times$10$^{-7}$ \citep{larsson07}.   One favored explanation for the surprisingly low \xotwo\ is gas-grain interactions: atomic oxygen depletes onto grain surfaces in cold clouds, is subsequently hydrogenated to water, which (if the grain temperature is sufficiently low) remains in the form of water ice on the grain surface \citep{bergin00, hollenbach09} leaving relatively little gas--phase oxygen to form \otwo, especially after that tied up as CO is considered. 

The {\it Herschel} Oxygen Project (HOP) is an Open Time Key Program using the HIFI instrument \citep{degraauw10} on board the {\it Herschel} Space Observatory\footnote{{\it Herschel} is an ESA mission with science instruments provided by European-led Principal Investigator consortia and with important participation from NASA.} \citep{pilbratt10}.  The goal of HOP was to carry out a survey of three rotational transitions of \otwo\ toward a broad sample of dense clouds, in which gas-phase {\otwo} abundance is predicted to be enhanced, including massive star forming regions, shocked regions, and photodissociation regions (PDRs).  Compared to {\it SWAS}, the noise temperature of HIFI at the 487 GHz \otwo\ frequency is $\simeq$ 28 times lower and the beam solid angle of  {\it Herschel} a factor $\simeq$ 33 times smaller.  These characteristics enabled \citeauthor{goldsmith11} (2011) to report the first multi-line detection of \otwo\ toward Orion \peakone\ with a beam-averaged column density of \Notwo\ = 6.5$\times$10$^{16}$ cm$^{-2}$ and a derived abundance of \xotwo $\sim$10$^{-6}$. 

Also as part of HOP, two {\otwo} transitions were detected towards the $\rho$\,Oph core and a fractional abundance \xotwo $\sim$5$\times$10$^{-8}$ was inferred \citep{liseau12}. A combination of the HIFI data with the earlier Odin detection at 119 GHz \citep{larsson07} suggested that the \otwo\ emission must be spatially extended.  \citeauthor{melnick12} (2012) reported an upper limit to $N$(\otwo) implying a face-on column density (the relevant quantity for PDR--produced \otwo) of less than 4$\times$10$^{15}$ cm$^{-2}$ toward the Orion Bar PDR. Toward the low-mass protostar NGC~1333~IRAS~4A, \citet{yildiz13} found an upper limit of \xotwo $\le$5.7$\times$10$^{-9}$. 

The derived \xotwo\ toward \peakone\ from the HIFI observations is still below the predictions of pure gas-phase chemistry, but much higher than the upper limits obtained in other sources. To explain the relatively high \xotwo\ toward \peakone, \citeauthor{goldsmith11} (2011) suggested two possible mechanisms: (1) thermal desorption of water ice from warmed dust grains, allowing subsequent {\otwo} formation in the gas-phase \citep[e.g.][]{wakelam05} and (2) enhancement of \xotwo\ in shocked gas \citep{kaufman10}. \peakone\ is the most strongly shocked position traced by vibrationally excited molecular hydrogen in the Orion Kleinmann-Low (KL) region, which is known for complicated massive star-forming activities, with interactions between outflows and the ambient material \citep[e.g.][]{bally11}. Therefore, with only one pointing direction, there is not enough information to pinpoint the source of emission in order to distinguish between the two scenarios. 

As seen in previous molecular line surveys toward the Orion~KL region, many molecular lines show complex line shapes, which can be attributed to several spatial components: the so-called Orion Hot Core at velocity with respect to the local standard of rest (\vlsr)  of 3--5~\kms\ , the Compact Ridge at \vlsr = 7--9~\kms\ , and the Plateau (low--velocity outflow) at \vlsr = 6--12~\kms\ \citep[e.g.][]{blake87}. The observed HIFI spectra toward \peakone\ show a line feature at 10--12~\kms\ in all three {\otwo} transitions at 487 GHz, 774 GHz, and 1121 GHz. An additional 5--6 \kms$\,$ feature is also seen in the 487 GHz observation. 

On the basis of their observations carried out with a 44\as\ beam at 487 GHz, \citeauthor{goldsmith11} (2011) suggested that the feature at 5--6~\kms\ could be an {\otwo} emission from the Hot Core. However, there could also be gas along the same line of sight having velocity 7--8 \kms characteristic of much of the Orion region, in which case the feature could also be methyl formate emission at this higher velocity. The 10--12~\kms\ velocity is unusual for molecular emission in this region, which is largely characterized by lower velocities. A number of high angular resolution studies suggest a small source west of Orion IRc2 (hereafter \peaka), approximately 23\as\ away from \peakone, with line emission in the 10 to 14~\kms\ range \citep[e.g.][]{masson1988,wright1996,goddi2011}. \peaka\ is the only source with narrow lines and molecular emission in range 10--12~\kms, and the line excitation condition also suggests that it could be warm enough for water desorption (T $>$ 100 K), which makes it a good candidate to be responsible for the observed {\otwo} lines. To investigate the true source of the emission and to explain the derived abundances, HIFI observations of the three transitions observed in \peakone\ were obtained towards \peaka.

In this paper, we detail the observations and the data reduction in Section 2. We present the observational results in Section 3. In Section 4, we discuss the line identification process and effects of beam filling. The implications of the origins of the {\otwo} emission are discussed in Section 5, and the summary of our results are presented in Section 6.  The formalism for the beam coupling calculations is given in the Appendix.

\section{Observations}

The observations toward \peaka\ of the \otwo\ transitions at 487 GHz and 774 GHz were carried out in 2012 April. The observations for the 1121 GHz transition were performed in 2012 August and September. The observed transitions, line frequencies, upper level energies (\Eu), observing dates, the OBSIDs, and the pointing offsets of the H and V beams are listed in Table~\ref{table:observations}. The  J2000 coordinates are $5^{\mathrm{h}}35^{\mathrm{m}}14\fs2$, $-5\degr22\arcmin31\farcs$ Figure~\ref{fig:co} shows the pointing positions of \peaka\ and \peakone. We used HIFI in dual beam switch (DBS) mode with the reference positions located 3$'$ on either side of the source. For each transition, eight local oscillator (LO) settings were used to allow sideband deconvolution. The integration time for each LO setting was 824 seconds for the 487 GHz and 774 GHz spectra, and 3477 seconds for the 1121 GHz spectrum.

The data were processed with the {\it Herschel} Interactive Processing Environment (HIPE) version 9.1 \citep{ott2010} and exported to CLASS, which is part of the IRAM GILDAS software package\footnote{http://www.iram.fr/IRAMFR/GILDAS}, for further analysis. Due to the higher noise level of the High Resolution Spectrometer (HRS) and its narrower bandwidth that prevents deconvolution, we include only the results obtained with the Wide Band Spectrometer (WBS) in this paper\footnote{The deconvolution of the double-sideband spectra requires the frequency range of interest to be covered with multiple, slightly shifted frequency settings \citep{comito2002}. Given the narrow bandwidth of the HRS, 235 MHz \citep{degraauw10}, as compared to 4 GHz for the WBS, the individual spectra do not overlap in frequency and the deconvolution cannot be carried out.}. The task DoDeconvolution is applied in HIPE to extract single-sideband spectra. The two linear polarizations (H and V) do not show any appreciable intensity differences and are added together, except that there is one corrupted frame in the V polarization from the OBSID 1342250406 spectra at 1121 GHz that has been removed. The two polarizations have small (1{\arcsec} to 3{\arcsec}) pointing offsets depending on the band and the observing date.  In \citet{goldsmith11}, the intensity differences were used with the knowledge of polarization offsets to suggest a position for the emitting source (displaced from \peakone\ in the direction of \peaka).  The lower signal-to-noise ratio of the present data and the additional uncertainty introduced by the greater line confusion here prevent us from using the two polarizations independently.

\section{Observational Results}

Figure~\ref{fig:o2} shows the baseline-subtracted spectra toward \peaka. The line parameters derived from the spectra are summarized in Table \ref{table:parameters}, which gives the integrated intensity ($I$), \vlsr, the line width ($\Delta V$), and the peak antenna temperature (\Ta). The values for $I$, \vlsr, $\Delta V$, and \Ta\ are determined from Gaussian fits to the lines. The main beam efficiency is 0.76, 0.75, and 0.64 at 487 GHz, 774 GHz, and 1121 GHz, respectively \citep{roelfsema12}. 

The line features at 487 GHz and 774 GHz are similar to the results from the \peakone\ observations with \vlsr\ 10$-$11 {\kms}, but the line at 1121 GHz is only a 3$\sigma$ tentative detection. The spectra show lines from many molecular species with significant line blending, which makes baseline fitting difficult. For the 487 GHz spectrum, the wing of a nearby strong SO$^+$ line with a peak intensity of $\sim$0.5~K at 487212.1 MHz was removed by fitting a second order polynomial baseline. For the 774 GHz spectrum, a third order polynomial baseline for the velocity interval  0--25~\kms\ was applied, avoiding a blend of dimethyl ether (CH$_3$OCH$_3$) lines around $\sim$773869 MHz with intensity $\sim$1~K and a CH$_3$OH line at 773892.54 MHz with a peak antenna temperature of about 5~K. In the 1121 GHz spectrum, there is a $\sim$0.3~K $^{13}$CH$_2$CHCN line at 18 \kms\ and an unknown line at about 13 \kms. To remove the contribution of the $^{13}$CH$_2$CHCN line to the \otwo\ line, we applied a third order polynomial baseline. The results for all \otwo\ line intensities are sensitive to  the baseline placement, and we assume a 20\% uncertainty for the integrated intensity resulting from the baseline fitting for the 487 and 774 GHz lines and a 0.02 K \kms\ uncertainty for the limit on the 1121 GHz \otwo\ line, based on different methods of baseline removal of the relatively strong nearby lines. Table \ref{table:uncertainties} summarizes the statistical uncertainties from the Gaussian fittings, the instrumental  calibration uncertainties, the baseline fitting uncertainties, and the combined uncertainties for the integrated intensities.


\section{Analysis}
\subsection{Line Identification}

We examined all molecular transitions found in the \verb1SPLATALOGUE1 catalog\footnote{http://www.splatalogue.net} falling within 5 \kms\ of each targeted \otwo\ line, in order to  confirm the O$_2$ detection and rule out interlopers, as discussed in \citeauthor{goldsmith11} (2011). For each species that has transitions within 5 \kms\ of our \otwo\ lines, we used the \verb1XCLASS1 program\footnote{https://www.astro.uni-koeln.de/projects/schilke/XCLASS} to model the spectra for transitions having upper level energy \Eu\ up to 1000~K. The XCLASS program includes entries from the JPL \citep{pickett98} and CDMS \citep{muller01,muller05} line catalogs and can provide a simultaneous fit for all the lines in the spectra under the assumption that the populations of rotational levels are described by local thermodynamic equilibrium (LTE) \citep{comito05, zernickel12, crockett14}. The parameters include the source size, the column density ($N$), the excitation temperature (\Tex), \vlsr, and $\Delta V$. This method is very efficient and accurate when dealing with line blending and multiple velocity components. We also successfully identified a few other molecular lines present in our spectra, and compare the characteristics with the detected O$_2$ lines.

\subsubsection{Methyl Formate}
Based on our modeling, we can rule out as possible contaminants most molecular species with transitions that lie near the \otwo\ lines.   For example, although methylamine (CH$_3$NH$_2$) has several transitions ranging from 487242 to 487253 MHz, no other transitions at different frequencies in the observed spectra were found, which suggests CH$_3$NH$_2$ cannot contribute much to the 487 GHz O$_2$ feature. The only problematic species is methyl formate (CH$_3$OCHO), for which the CH$_3$OCHO 40$_{6,34}$--39$_{6,33}$ transition at 487252 MHz, with an upper level energy (\Eu) of 705 K, interferes with the feature that could be \otwo\ at the Hot Core velocity 5--6~\kms. All CH$_3$OCHO transitions up to \Eu=1000~K are identified across the spectra, which confirms the detection and the contribution of CH$_3$OCHO at this frequency. To estimate the line intensities for this blended line feature, we focus on fitting relatively isolated, unblended transitions with similar \Eu\ ($\sim$700 K). Figures \ref{fig:mf_peakA} and \ref{fig:mf_peak1} show comparisons between modeled and observed CH$_3$OCHO lines toward both observed positions. 

Table \ref{table:mf} lists the transitions, line frequencies, A--coefficients, and \Eu\ from the JPL catalog \citep{pickett98} for the CH$_3$OCHO lines modeled in Figures \ref{fig:o2}, \ref{fig:mf_peakA}, and \ref{fig:mf_peak1}. The best-fit parameters assuming a 5\as\ FWHM source size are: $N$=1.4$\times$10$^{17}$ cm$^{-2}$, \Tex=125~K, $\Delta V$=2.3~\kms, \vlsr=7.7~\kms. The same analysis was carried out with data toward \peakone, where the CH$_3$OCHO lines are much weaker. The fitted parameters are:  $N$=1.0$\times$10$^{17}$ cm$^{-2}$, \Tex=100~K, $\Delta V$=2.3~\kms, \vlsr=7.7~\kms, again assuming a 5\as\ source size. In principle, we can compare the modeled CH$_3$OCHO lines in the 774 and 1121 GHz bands to further constrain the source size and position, but with less sensitivity, as the lines with \Eu\ $\sim$700 K in these bands are too weak. Figure~\ref{fig:residual} shows the comparison of the modeled CH$_3$OCHO lines and the observed \otwo\ lines at 487 GHz toward both \peaka\ and \peakone. The area of the residual (red lines) in the range of 0--11 \kms\ is 1.0$\times$10$^{-1}$~K \kms\ and 1.1$\times$10$^{-1}$ K \kms\ toward \peaka\ and \peakone, respectively. If all or part of this feature is \otwo\ emission from the Hot Core region, we would expect to see stronger lines in both 774 and 1121 GHz toward the \peaka\ position. Since we see a 5--6~\kms\ feature only in the 487 GHz data, the contribution of \otwo\ emission from the Hot Core is likely insignificant.

\subsubsection{Sulfur Monoxide}
\label{SO}
We identified lines from CH$_3$OH, C$_2$H$_5$OH, CH$_3$OCH$_3$, CH$_3$OCHO, HNCO, C$^{17}$O, H$_2$CO, SO, SO$_2$, H$_2$CCO, NS, OCS, and H$_2$CS in our {\it Herschel} observations. All of the lines are relatively narrow and peak at a velocity of 7--8 \kms, with the exception of the emission from SO and SO$_2$, which is dominated by a wide line component at 6~\kms\ from the Orion plateau. 

As described in previous sections, the 11~\kms\ velocity of the narrow-line \otwo\ emission is rather unusual in this region. Among all molecules identified, we found that SO is the only species observed with {\it Herschel} having a narrow 11~\kms\ line component similar to the \otwo\ lines. The chemical relationship between these two species has been discussed by \citet{nilsson00}, who show that in standard gas--phase chemistry models, the abundances of both SO and \otwo\ become appreciable only at ``late'' times when most carbon is locked up in CO.  The SO lines appear to be blends of a broad feature centered at about 6 \kms\ and a narrower feature centered at about 11 \kms. Figure \ref{fig:so} shows the comparison of the SO lines toward \peaka\ and \peakone\ in both the 487 GHz and the 774 GHz bands.  Relative to the 6 \kms\ broad feature, the 11~\kms\ component is significantly more enhanced toward \peakone.  This is consistent with SO abundance being enhanced in a shock \citep{espluges13}, and that the emission having 11 \kms\ velocity is a signature of the postshock gas.  NO is another chemically--related species, and we previously \citep{goldsmith11} reported observations of NO with the Caltech Submillimeter Observatory (CSO) displaying a narrow component at velocity 10.4~\kms\ having width 3.1~\kms.   A shock producing  OH would then enable OH + N $\rightarrow$\ NO + H (analogous to OH + O $\rightarrow$\ \otwo\ + H), assuming that atomic nitrogen is present in the gas phase.  The existing data are not adequate  to determine the degree of positional agreement with the \otwo\ emission. 

\subsection{Beam filling and source offset effects}

In the previous study of \otwo\ in Orion \citep{goldsmith11}, two candidate sources were proposed to explain the enhanced \otwo\ abundance. The first source is a maximum of  the vibrationally excited H$_2$ emission, generally attributed to a shock, which has one of its maxima at the position observed, denoted ``\peakone".  The second source is ``\peaka", which is a subregion of the Orion Hot Core. These two sources are separated by 23\arcsec, which corresponds to 0.05 pc at a distance of 420 pc \citep{menten2007}.
This is comparable to the {\it Herschel} FWHM beam sizes, which range from 44\arcsec\ at 487 GHz to 19\arcsec\ at 1121 GHz.
A low--velocity shock could enhance the \otwo\ abundance, making \peakone\ a plausible location for the \otwo\ emission, but there has been no clear kinematic association of that position with the relatively narrow emission centered at a velocity of $\simeq$11\kms.  
\peaka, being close to the luminous sources in Orion, might well have dust sufficiently heated to desorb water ice mantles, leading to reestablishment of gas--phase chemistry, in which the fractional abundance of \otwo\ is expected to be in excess of 10$^{-6}$ after sufficient time has elapsed (see Goldsmith et al. 2011).

For the new data discussed above, the beam pointing direction is essentially that of the \peaka\ position, albeit with the caveat that there are small pointing offsets between the two polarizations (as discussed above).  
If we assume that there is a single source having an angular size comparable to, or smaller than the separation between the two observed positions, there should be an appreciable difference between the two observed intensities.  
In particular, if the source were located at the \peaka\ position, we would expect the line intensities at that position to be significantly greater than at the \peakone\ position. 
From the results given in Table \ref{table:pkApk1_relative}, this is clearly {\it not} the case. Rather, the results suggest that the emitting region must be located relatively closer to the \peakone\ position than to \peaka.

The formalism for calculating the coupling of the beams at the two observed positions to a single source is given in the Appendix.  If the source is located on the line connecting \peaka\ and \peakone\, and their angular separation is $\theta_{\rm sep}$, then the ratio of the integrated intensities ($I = \int T_{\rm A}dv$)  can be written as depending only on the offset of the source from \peakone\ ($\theta_{01}$) and the convolved Gaussian width of the beam and the source ($\theta_{co}$; see equation \ref{theta_co} in the Appendix) as
\begin{equation}
\frac{I(obs~A)}{I(obs~1)} = \frac{exp[-((\theta_{\rm sep} - \theta_{01})/\theta_{\rm co})^2]}{exp[-(\theta_{01}/\theta_{\rm co})^2]} \lp
\end{equation}
Figure~\ref{fig:pkApk1_rel} shows the results for three source sizes (relative to the source separation and offsets of interest): point--like (2.5\as\ FWHM), modest (10\as\ FWHM), and significantly extended (20\as\ FWHM).  
The source size does not have a dramatic effect on the relative integrated intensities.
Rather, it is primarily the offset that determines the observed ratio.
The separation between \peaka\ and \peakone\ is 23\as, so that for an offset equal to half this value, the intensities at the two observed positions will necessarily be equal.  
As the offset from \peakone\ decreases from 11.5\as\ with the source being closer to \peakone, the \peaka~/\peakone\ ratio decreases.

An acceptable solution should have ratios for all three lines consistent with the uncertainties given in Table \ref{table:pkApk1_relative}.
With this restriction, we can only establish a range of source locations.
The 487 GHz data (with equal intensities) indicate that the offset is greater than 6\as\ for a small source (2.5\as\ or 10\as\ FWHM) and greater than 5\as\ for a 20\as\ FWHM source.
The 774 GHz data restrict the offset to between 6\as\ and 8\as\ for a small source and between 5\as\ and 7\as\ for a 20\as\ source.
The 1121 GHz data do not place a significant constraint on the offset due to the relatively low signal to noise ratio.
For source size $\leq$ 10\as, an offset of 6\as\ to 8\as\ from \peakone\ is consistent with the data for the three transitions.  
For a 20\as\ source size, there is a slightly different but similar--sized range of offsets, 5\as\ $\leq \theta_{01} \leq$ 7\as, that is consistent with all of the observations.
Given the uncertainties in the data, we feel that the observed intensity ratios rule out \peaka\ as the source of the \otwo\ emission, and suggest that it is closer to \peakone, displaced by somewhat less than half the distance between \peakone\ and  \peaka.  

There is, in fact, no restriction that the source be located on the line between \peaka\ and \peakone.  
The ratio of the observed intensities for a given transition at the two positions defines the ratio of the distance of the source to the two observed positions. 
The locus of source positions that produce a given observed ratio less than unity is a circle.  It intersects the line connecting \peaka\ and \peakone\ at the position defined by $\theta_{01}$, and also at a second position on the opposite side of \peakone\ from \peaka.  
These are generally in the region of the prominent vibrationally excited H$_2$ emission seen in Figure 1 of \citet{bally11}.  
A ratio of unity translates to a line perpendicular to that connecting \peakone\ and \peaka, and passing through its midpoint.
While source positions not located between \peaka\ and \peakone\ are possible, they share the characteristic that the absolute distances to the two observed positions are greater than that for the solution on the line connecting the sources.
For this reason, the offset coupling factors (equation \ref{Ta_1}) are smaller and so a greater column density of \otwo\ is required to produce the observed antenna temperatures.
As discussed in \S \ref{shock}, reproducing the observed column density of \otwo\ is a challenge, and a larger column density makes it even more difficult.
Thus, while we cannot eliminate any of these positions, a source 6\as--8\as\ from \peakone\ in the direction of \peaka\ is the best explanation for the \otwo\ data obtained to date.

\subsection{\otwo\ Column Density}
We used XCLASS to model the observed \otwo\ emission toward the \peakone\ and \peaka\ positions to determine the \otwo\ column density. 
The calculations assume that the telescope beam is centered on the source; therefore, the offset-coupling correction (Sect. 4.2) has to be applied separately before computing model intensities toward the \peakone\ and \peaka\ positions. For a given source size, we have varied the input column density and kinetic temperature to obtain the best fit to the six observed line intensities. The results, shown in Table \ref{table:densities}, indicate that a source size of $\sim9$\as\ provides the best fit to the data. The observations imply a low kinetic temperature of $\sim30$ K, and the resulting peak \otwo\ column density in a pencil beam is $\sim1\times10^{18}$ \cmc. We have verified that a source location $\sim7$\as\ from \peakone, as given by the analytic formulae, indeed provides the best fit to the data.

\subsection{Limit on \otwo\ in the Hot Core}

We previously showed that methyl formate alone is unlikely to explain all of the 5--6 \kms\ feature seen in the 487 GHz spectrum of \peaka\ (see Figure 5). We can address whether this feature might be in part due to \otwo\ from the Hot Core. For temperatures greater than 40 K the 774 GHz line of \otwo\ is stronger than the line at 487 GHz \citep{goldsmith11}. Therefore for the  high temperatures associated with the Hot Core, the 774 GHz line should be much more sensitive to the presence of \otwo\ than the line at 487 GHz. The 774 GHz spectrum is presented in Figure 2 and shows little evidence for \otwo\ emission at 5--6 \kms. Therefore we believe it is unlikely that much of the residual present after the methyl formate fit for the 487.252 GHz line can be due to \otwo\ emission.

We can use the 774 GHz line to set a limit to the column density of \otwo\ in the Hot Core. A 1$\sigma$ upper limit to the integrated intensity of \otwo\ in the velocity range of 0-10\kms\ is 0.013 K \kms. A significant correction needs to be made to account for the fact that the Hot Core is offset from our \peaka\ pointing direction and that the source is smaller than the Herschel beam size at 774 GHz. Based on the observations of \citeauthor{plambeck87} (1987) and \citeauthor{masson1988} (1988), we estimate that the Hot Core has a Gaussian source size of 8\as\ and that the position of the Hot Core is offset relative to the \peaka\ pointing by 4.7\as.  Using equations 7-11 presented in the Appendix, and assuming a small source offset from the telescope pointing, we estimate the correction factor to be 0.054. Therefore the corrected 1$\sigma$ upper limit on the integrated intensity of the 774 GHz line of \otwo\ in the Hot Core is 0.24 K \kms.

A recent paper  \citep{neill13} tabulates the properties of the Hot Core. We assumed a density of 10$^7$ cm$^{-3}$ and a temperature between 150 and 300 K. We used RADEX \citep{vandertak07} with the collision rates of \citet{lique10} to compute the integrated intensity of \otwo\ for a fixed H$_2$ density of 10$^7$ cm$^{-3}$, together with different \otwo\ column densities and temperatures.  We find that for 150 K, the one-sigma limit on the \otwo\ column density is 1.2$\times$10$^{17}$ \cmc\ and for 300 K, the upper limit is 1.9$\times$10$^{17}$ \cmc. The total H$_2$ column density in the Hot Core is estimated to be 1.3$\times$10$^{24}$ \cmc\ \citep{favre11}.  Therefore the 1$\sigma$ upper limit on the \otwo\ abundance relative to that of H$_2$ is 0.9-1.5$\times$10$^{-7}$. Since the Hot Core is small and offset from \peaka, our data do not put a very stringent limit on the relative \otwo\ abundance.

\section{Discussion}

A previous paper on O$_2$ emission from the central region of the Orion molecular cloud \citep{goldsmith11} suggested two possible explanations for a localized enhanced abundance of molecular oxygen, which were (1) desorption of grain--surface ice mantles from warm dust surrounding an embedded heating source, and (2) production of molecular oxygen in a low--velocity shock.  The former model suggests identification with the \peaka\ region of the Hot Core, possibly heated by IRc7, supported by the rather unusual velocity of $\simeq$ 11 \kms, found only in this region.  The latter suggests that the emission is associated with \peakone, the location of strongest vibrationally--excited H$_2$ emission, which is thought to be a consequence of a shock.  

The present observations weigh strongly against the emission being centered near \peaka.  Rather, the best--fit parameters for the source of the emission place it nearer \peakone, albeit somewhat displaced in the direction of \peaka.  We thus want to consider in more detail other evidence about the geometry of the shock in this region and the effects it may have produced, as well as the production of the observed O$_2$ by a shock. \citet{goldsmith11} summarize the information about the H$_2$ emission of relevance to the shock--production scenario for O$_2$.  \otwo\ production in a 25 \kms\ shock was considered by \citet{draine83}, and for a range of shock velocities by \citet{kaufman10} and \citet{turner12}.

\subsection{H$_2$O maser emission and O$_2$}

An interesting and significant observational result that bears on the origin of the O$_2$ emission is the recent discovery of a maser feature in the 5$_{32}$-4$_{41}$ transition of H$_2$O at 620.701 GHz by \citet{neufeld13}.  In a region situated 20\as--50\as\ to the north and 0\as--20\as\ to the east of Orion-KL, there is a narrow feature at 12 \kms\ velocity with a line width $\leq$ 2 \kms , which stands out above the $\geq$ 5 \kms--wide thermal emission.  The similarity of the water maser feature's velocity with that of the O$_2$ is suggestive, given that water masers are often associated with shocks.

In exploring this connection, it is important to recognize that a maser will have its maximum gain in a direction along which there is maximum velocity coherence (or minimum velocity gradient).  The fact that we see the 621 GHz water maser emission is consistent with the line of sight being essentially  perpendicular to the direction along which the shock is propagating, since  the water molecules would inevitably be spread out in velocity by the passage of the shock.  It is also consistent with the velocity of the emission, which is shifted by only 3--4 \kms\ relative to the velocity of the ambient cloud.  Thus, since the velocity of a shock that could produce sufficient excitation for this maser transition as well as significant \otwo\ \citep{kaufman10} would be several times greater than this, the velocity vector of the shock is not too far from the plane of the sky.  

Given this orientation, it is likely we are viewing the shock almost perpendicular to its velocity vector.  Consequently, it is not the column density (of \otwo) parallel to the shock velocity that is the critical quantity, but rather the fractional abundance of \otwo\ behind the shock multiplied by the H$_2$ column density of the shocked region along the line--of--sight, which together yield N(\otwo) perpendicular to the shock velocity vector.  We thus focus on the fractional abundance of \otwo\ in our discussion of shocked models that follows, and then discuss the required line--of--sight dimension of the region that is required to reproduce the observed \otwo\ column density.


\subsection{\otwo\ Emission in Orion from Shocked Gas}
\subsubsection{Modeling the shock}
\label{shock}

In order to assess whether a shock can provide the observed column density of \otwo, and to get an idea of what constraints on shock and environmental parameters are provided by our observations, we have employed an efficient parameterized C--shock code \citep{jimenez08} coupled with the time--dependent gas--grain chemical code UCL$_{-}$CHEM \citep{viti04}. Details of the coupled code can be found in \citet{viti11}.   The parameterized model includes a magnetic field that varies with density according to equation 10 of \citet{draine83}.  \citet{kaufman10} and \citet{turner12}  found previously that low--velocity shocks are most efficient at producing O$_2$, because a gas temperature of $\simeq$ 1000 K allows acceleration of OH production via H$_2$ + O $\rightarrow$ OH + H (followed by OH + O $\rightarrow$ O$_2$ + H), but not by so much that the supply of O is exhausted.  High--velocity shocks allow the back reaction O$_2$ + H $\rightarrow$ OH + H, which has a barrier of 8750 K.  We adopt a shock velocity of 12 \kms, the velocity found by \citet{kaufman10} and \citet{turner12} to produce the maximum column density of \otwo.

We would like to draw the reader's attention to one of the limitations of our model.  This is, that the parametrization of C-type shocks employed in the time dependent gas-grain code UCL$_{-}$CHEM does not include the explicit time dependence of the physical parameters of the shock, in that the partial time derivatives are set to zero. The main limitation of this approach is that, under certain circumstances, the physical parameters of the shock may change dynamically over time scales comparable to or shorter than the flow time through the shock. Chemically, it may be that the assumption of steadiness for our C-type shock may not be adequate to describe accurately the time evolution of the sputtering of the grains.  Nevertheless, as already discussed, a low sputtering efficiency is sufficient to produce a high abundance of \otwo\ under the conditions that we have considered.  \citet{jimenez08} compares results of the parameterized model with more complete shock models, and the limitations of steady-state C-type shocks are  discussed in \cite{chieze98}, \citet{lesaffre04a}, and \citet{lesaffre04b}, and we refer the reader to these papers for details.  

The modeling used here starts with a phase during which an isolated clump evolves in time, with density increasing from $\simeq$ 100 \cmv\ to final density equal to 2$\times$10$^4$ \cmv. During this Phase I,  the kinetic temperature is 10 K (characteristic of a quiescent, dark cloud rather than Orion) chosen to highlight the effects of grain--surface depletion and chemistry,  as discussed in \citet{viti11}.  We include a cosmic ray ionization rate $\zeta$ = 10$^{-17}$ s$^{-1}$; this is significantly lower than suggested by \citet{indriolo07} for regions of low column density.  But the large extinction and column density in the core of Orion suggest that the lower value may be appropriate.  The cosmic ray ionization rate does not affect the shock production of \otwo\ significantly. It does determine the timescale for reestablishment of ``standard'' gas--phase chemistry in the postshock gas, as seen in Figure 8 of \citet{goldsmith11}.

To study the effects of depletion, we have considered two durations for this phase of the evolution of the condensation:  a ``short'' Phase I, lasting lasting 5.3$\times$10$^6$ yr and a ``long'' Phase I lasting 7$\times$10$^6$ yr.  Despite the relatively small  difference in elapsed time, the exponential behavior of depletion onto grain surfaces makes the gas--phase abundances of key species quite different at the end of the two Phase I durations.  However, the effects on the ultimate evolution of the \otwo\ abundance are relatively modest, so  our conclusions are not very sensitive to the pre--history of the shocked gas.  

Based on the fitting shown in Table \ref{table:densities}, we adopt for the preshock material a condensation having size in the plane of the sky equal to 9\as, corresponding to a linear dimension of 0.018 pc at the distance of Orion.  This condensation is assumed to be embedded in less dense material having a visual extinction of 2 mag.  The exact value is not critical inasmuch as the significant effect of this material is to attenuate the UV radiation field, which is not well known.  In Phase II, we assume an incident UV flux a factor $\chi$ times the standard Draine value \citep[2.1$\times$10$^{-4}$ erg\,cm$^{-2}$\,s$^{-1}$\,sr$^{-1}$,][]{draine78} incident on one face of a plane--parallel  region into which the shock is propagating.  We have considered $\chi$ = 1, 10, and 100.

Phase II includes the passage of the shock; the gas temperature during the shock is calculated by the shock model, but we assume the preshock temperature to be fixed at 50 K until the shock heating from ion--neutral velocity difference becomes dominant.  This is arbitrary and has no significant effect on the chemistry during the less than 100 yr period before the temperature rises appreciably.  After the passage of the shock, the gas cools but we constrain the minimum temperature following passage of the shock to be either 50~K or 100~K.  These two values of the postshock minimum temperature  produce essentially identical results, and there is no issue with the implied temperature of 30~K from the \otwo\ observations.  The relatively low temperature indicated by the \otwo\ is consistent with it not being within the Hot Core or Peak A, which are considerably warmer, and also implies that the \otwo\ observed must be produced well downstream from the peak of the shock heating, a result that is entirely consistent with the model results presented in what follows.

%
%
The evolution of the density and temperature in the postshock gas are shown in Figure \ref{fig:dens-temp} for the case of a model having minimum temperature of 100 K.  It takes about 150 years after the shock reaches a particular point in the region for the water on the grain mantles to be sputtered by the shock.   The peak gas temperature in the shock is 600 K.  Redepletion onto grain surfaces is included in the postshock evolution, but at the temperatures considered here, it is inefficient and unimportant at the time scales considered here.  


A possibly significant effect of the shock is the sputtering of icy grain mantles.  This is obviously of greater potential importance for the long Phase I, in which a large fraction of the initial oxygen reservoir has been depleted onto grains and hydrogenated to ice, than in the short Phase II.  There has been significant evolution in the calculations of the minimum shock velocity required for efficient sputtering.  \citet{draine83} suggested that 20 \kms\ is required to get 10\% sputtering of a water ice mantle, but \cite{flower94} suggested that shock velocities as low as 10 \kms\ could produce significant mantle sputtering.  \citet{draine95} pointed out that the threshold for sputtering should be a factor of 4 higher than used by \citet{flower94} and \citet{flower10} concluded that sputtering would be significant for shock velocities as low as 15 \kms\ but unimportant at 10 \kms.   Given this uncertainty, as well as that of the appropriate shock velocity, we have adopted a 10 \% sputtering efficiency at the 12 \kms\ shock velocity.  We have explored the effect of lower and higher sputtering efficiencies, which are discussed after we present the results of our ``standard'' model.

Figures~\ref{1uvshort} and~\ref{1uvlong} show the evolution in Phase II of the fractional abundances of selected species for two models differing in the durations of Phase  I.  
Both models have been run with an external UV field with $\chi$ = 1.  
For the short Phase I, we see that the {\it initial} gas--phase abundance of atomic oxygen is large, although that of water is small, due to the grain--surface depletion at the 10 K temperature of the cloud in Phase I, which determines the conditions at the start of the shock passage in Phase II.  For the long Phase I, the gas phase abundance of atomic oxygen is lower by more than a factor of 100, and those of other molecular species are also significantly reduced. Table~\ref{table:reactions} lists some of the dominant ($>$40\% fractional contribution to the total rate) reactions leading to the formation and destruction of O$_2$ (and its parent and daughter species) in different time intervals within Phase II for the Model in Figure~\ref{1uvshort}.  Note that for some periods of time there may not be a single dominant reaction for the formation of a particular species. 

The model outputs shown in Figures~\ref{1uvshort} and~\ref{1uvlong} indicate that molecular oxygen is efficiently produced for a limited period of time. The shock models indicate that the behavior of SO and NO (not shown) is similar to that of O$_2$ in that they are abundant (and deficient) at similar times (as suggested in the discussion in Section~\ref{SO} above). Hence if the \otwo\ shock--production scenario is correct, we predict that these species should be abundant at the same time as O$_2$. Of course there are many caveats that one has to consider, most importantly that these species may be abundant in scenarios other than in shocks. More detailed modeling of O$_2$ is beyond the scope of this paper, but should be carried out in order to confirm our picture and make predictions of the relative distribution of the different species.

In both cases, the \otwo\ abundance increases by several orders of magnitude following the passage of the shock with associated temperature rise (at time = 100 yr), but X(O$_2$) continues to rise for up to 10$^4$ yr following the shock passage, due to the gradual conversion of gas--phase atomic oxygen into molecular oxygen, with fractional abundance X(\otwo) reaching a maximum value $\simeq$ 5$\times$10$^{-5}$ for short Phase I and 2$\times$10$^{-5}$ for long Phase I.  Subsequently, the abundances of both molecular oxygen and water decline and that of atomic oxygen increases due to photodestruction of the molecular species. 

In Figure \ref{0uv} we show behavior for a short Phase I, but with no UV radiation present.  The abundances immediately following the passage of the shock are not very different from those in  Figure~\ref{1uvshort}, but the \otwo\ abundance  increases only to $\simeq$ 10$^{-6}$ during the period of shock heating, and remains at this level for a considerable time.  Only with the very slow destruction of H$_2$O (by cosmic rays and other destruction channels) do the OH and \otwo\ abundances start to increase; this requires $\simeq$ 3$\times$10$^4$ yr.  At very long times ($\geq$ 10$^5$ yr) after passage of the shock front, the \otwo\ abundance does exceed 10$^{-5}$.  This is the same behavior as seen in Figure 8 of \citet{goldsmith11}.  This long timescale is likely not of general relevance for regions in which massive stars are forming.  
 
Figure \ref{10uv} shows the results for a UV radiation field with $\chi$ = 10 and a short Phase I.  Here, the O$_2$ fractional abundance rises much more rapidly than in Figure~\ref{1uvshort}, and attains a value, X(O$_2$) $\simeq$ 3$\times$10$^{-6}$ only 300 years after the passage of the shock front, and  rises to $\simeq$ 4$\times$10$^{-5}$ 1000 years after the passage of the shock front.  After remaining only briefly at this value, the abundances of molecular oxygen and of water decline quite rapidly; the former is back to $\leq$ 1$\times$10$^{-6}$ only 10$^4$ yr after the passage of the shock front.  The asymptotic value X(O$_2$) $\simeq$ 1$\times$10$^{-8}$ is similar to that for the lower UV case, but is reached after only 10$^5$ yr, a factor of 10 sooner than for the lower UV field.  For a radiation field yet 10 times stronger ($\chi$ = 100), we find that the maximum abundance of O$_2$ is reduced to $\simeq$ 10$^{-5}$, and that this level lasts only for 500 years after the passage of the shock front.  The postshock asymptotic fractional abundance is X(O$_2$) = 10$^{-9}$.


In Figure~\ref{1uv10^5}, we show results of a model with a preshock density of 10$^5$ cm$^{-3}$ for a UV radiation field with $\chi$ = 1 and short Phase I.  Due to the higher density, the freeze out is more efficient  so the gas--phase O abundance is reduced.  Along with this, the low adopted sputtering efficiency in Phase II limits the rate of return to the gas--phase.  As in the case of zero UV radiation field, we see that the O$_2$ fractional abundance never reaches even 10$^{-5}$ in the ``postshock'' phase due to the relatively low efficiency of photodissociation resulting from the order of magnitude increase in the column density.  Most of the material in the clump is effectively shielded from the UV and the abundance of gas--phase water is large, and that of molecular oxygen is small, as in Figure \ref{0uv}. It is not until very late times, when oxygen is efficiently formed by several neutral--neutral reactions, that the abundance of O$_2$ rises to a value $\simeq$ 10$^{-5}$.  The reaction OH + OH contributes about $\sim$ 30\% to the formation of oxygen; at these late times OH is formed by several channels including the reaction of He$^+$ with H$_2$O.

We have carried out a limited investigation of the effect of shock velocity on the production of \otwo\ by running a model with a shock velocity of 15 \kms\ and conditions otherwise identical to those in Figure~\ref{1uvshort}.  The peak X(\otwo) drops slightly, from 5$\times$10$^{-5}$ to 3$\times$10$^{-5}$.  This is consistent with the no--UV behavior found by \citet{turner12} and no initial oxygen depletion.  However, if the sputtering efficiency were a rapidly increasing function of the shock velocity, the maximum \otwo\ abundance could well be achieved for somewhat higher shock velocities than the 12 \kms\ adopted here.

We have explored the effect of the sputtering efficiency on the production of \otwo\ in the shock.  For UV of $\chi$ = 1, if the sputtering efficiency were 100\% due either to the physics of the process or a higher shock velocity, then the maximum fractional abundance of \otwo\ would reach 10$^{-4}$ for both short and long Phase I, a factor of 2 and 5 higher, respectively,  than found with the 10\% sputtering efficiency models.  The same maximum X(\otwo) is achieved for a flux $\chi$ = 10, a factor of 3 higher than seen in Figure~\ref{10uv} for 10\% sputtering efficiency.  For no UV flux present, the sputtering efficiency has minimal effect on the \otwo\ abundance. If we eliminate grain mantle desorption  (by sputtering or other processes) entirely, X(\otwo) reaches 3$\times$10$^{-5}$ for a short Phase I and UV flux of $\chi$ = 1, compared to 5$\times$10$^{-5}$ for the 10\% sputtering efficiency presented in Figure~\ref{1uvshort}.  Particularly if the initial depletion of atomic oxygen is not severe, X(\otwo) as high as few$\times$10$^{-5}$ can be produced by the shock even in the absence of  grain mantle desorption.

\subsubsection{The source of UV radiation}

From the above we conclude that, as long as a flux of UV radiation with $\chi$ at least 1 is present, a large fractional abundance of \otwo\ can be produced by the shock, irrespective of the preshock atomic oxygen abundance.  A modest amount of UV is thus helpful in forming molecular oxygen in the context of a shock, with increasing UV shortening the time until maximum X(\otwo) is reached.  However, too much UV restricts the maximum fractional abundance of O$_2$ and also reduces the duration of the period in which X(O$_2$) is enhanced.  The exact values will depend on the external radiation field and the extinction both within and external to the region being shocked. The next question then is:  what is the source of the UV radiation that enables a sufficiently large \otwo\ abundance to reproduce the observations?

The Trapezium stars are a strong source of UV radiation; they affect not only their immediate environment, but also determine the large--scale distribution of [\CII] and [\OI] emission in Orion \citep{stacey93, herrmann97}.
The extended PDR on the front surface of the Orion molecular cloud in this picture is produced by UV radiation that travels from the Trapezium stars (located in front of the cloud surface as viewed from the Earth) attenuated only by geometric dilution, before entering the cloud and being absorbed.  
The \otwo\ source, as well as \peakone, are so close to the Trapezium that the UV radiation field could be attenuated by a factor approaching 10$^5$ and still be sufficient to produce the UV field employed in our irradiated shock models.  
The visual extinction to \peakone, based on the 2.12 $\micron$\ extinction of $\simeq$ 1 mag \citep{rosenthal00} and A$_V$/A$_K$  = 8.9 \citep{rieke85} is $\simeq$ 9 mag, corresponding to a visible--wavelength attenuation by a factor $\simeq$ 10$^4$.  
Since we do not know the distance of the region responsible for \otwo\ emission along the line of sight relative to the front side of the Orion cloud, the attenuation of UV radiation impinging on the shocked condensation must be regarded as highly uncertain (especially considering the effects of possible clumpiness), so that significant UV from the Trapezium cannot be ruled out.

A second source of UV radiation is higher velocity shocks that are known to be present in the region.  
Fast shocks (velocity greater than 50--80 \kms) produce significant amounts of UV radiation, particularly Lyman\,$\alpha$, which is particularly effective at dissociating H$_2$O \citep{neufeld89}.
\citet{kaufman10} proposed a combination of J--shocks and C--shocks to explain the emission lines observed from IC443, while \citet{vankempen09} pointed out the importance of heating by UV from shocks in the HH46 outflow.
In modeling the H$_2$ emission from Orion \peakone, \citet{rosenthal00} require excitation temperatures between $\simeq$ 600 K and $\simeq$3000 K.  These authors conclude that a combination of slow and fast shocks is required; the fast shocks or alternatively the transient J--shocks they consider may both be sources of UV radiation.
The extinction between \peakone\ and the \otwo--producing clump is also not well known, but it is plausible that high--velocity shocks could provide the UV flux for the models considered above.

\subsubsection{Shock production of \otwo\ in Orion and the \otwo\ abundance in molecular clouds}

The models we have run suggest that a clump with preshock density 2$\times$10$^4$ \cmv\ and H$_2$ column density $\simeq$ 4$\times$10$^{21}$~\cmc\ (such as found in the central portion of Orion in studies of various tracers summarized by \citet{goldsmith11}), after passage of a 12 \kms\ shock, can emerge as a postshock condensation with density 3.6$\times$ higher, but with an \otwo\ fractional abundance 3--5$\times$10$^{-5}$, if the UV radiation field present is characterized by 1 $\leq$ $\chi$ $\leq$ 10.     Thus, the postshock \otwo\ volume density is $\simeq$ 2~\cmv.  In order to provide the \otwo\ column density observed, 1.1$\times$10$^{18}$~\cmc, the line--of--sight dimension has to be 5$\times$10$^{17}$ cm, which is a factor $\simeq$ 9 greater than the fitted source size in the plane of the sky.  Thus, the geometry is not unreasonable (although perhaps not common), and is consistent with such an asymmetric region having maximum water maser gain along its larger dimension, which is along the line of sight and perpendicular to the velocity gradient produced by the shock.

In all of these models, the enhanced abundance of molecular oxygen is limited in time (the models refer to a fixed point in space as a function of time). Thus, the total column density of \otwo\ measured parallel to the shock velocity is also reasonably well-defined, especially in the cases with nonzero UV field.  We find for example that the total \otwo\ column density varies from $\simeq$ 10$^{17}$~\cmc\ for the $\chi$ = 1 case to $\simeq$ 10$^{16}$~\cmc\ for $\chi$ = 10, and a factor of 10 lower yet for $\chi$ = 100.  The column density for the $\chi$ = 1 case is consistent with the maximum \otwo\ column density predicted by the \citet{kaufman10} and \citet{turner12} models of shocks incident on regions with no preshock grain surface depletion and with no UV flux present.  

The crossing time for a 12\,\kms\ shock and region $\simeq$ 6$\times$10$^{16}$ cm in size is $\simeq$ 5$\times$10$^{10}$ s, or  $\simeq$  10$^3$ yr.  This is comparable to the duration of the elevated X(\otwo), so it is plausible that the entire postshock clump has an enhanced abundance of molecular oxygen.  But the short duration of the enhancement phase may well explain why none of the central regions of other massive star forming molecular clouds, most of which have undoubtedly experienced shocks as witnessed by prominent molecular outflows, has detectable \otwo\ emission. It may well be that it is the very recent activity in the central region of the Orion molecular cloud \citep[e.g.][]{bally11, goddi11}, as well as its proximity to the Earth, that lets us witness this dramatic but highly transient feature in its molecular composition.  We have not yet investigated all combinations of UV field, preshock density, shock speed, and clump size to find the very shortest time in which X(\otwo) can be elevated to a value approaching 10$^{-4}$, as needed to have  reasonable geometry.  The most rapid model studied (Figure~\ref{10uv}) has X(\otwo) reaching $\sim$ 2$\times$10$^{-6}$ only 200 years after passage of the shock front and remaining above 3$\times$10$^{-5}$ for $\simeq$ 3000 yr.    The timescale for shock production of \otwo\ is thus not inconsistent with the accurately--determined upper limit of 720 yr for the time since ejection of material and production of shocks in the center of Orion found by \citet{nissen12}.

The explanation for the size and precise location of the source of the \otwo\ emission is not yet apparent. 
Our modeling suggests that if a preexisting clump  with sufficiently high column density were present, the passage of a shock could dramatically raise X(\otwo) and produce the measured \otwo\ column density, as well as the water maser emission.  
Other species such as SO that in the shock modeling follow the time--dependence of \otwo, are seen to be enhanced at the same velocity.  
The fact that the peak \otwo\ emission is not coincident with the maximum H$_2$ emission associated with \peakone\ (which is itself quite extended and very complex in structure)  may be a result of the slower shock that maximally enhances X(\otwo), compared to that for producing the excited H$_2$.  
Calculations covering a range of shock velocities confirm that the much higher velocity shocks produce considerably less \otwo\ \citep{kaufman10,turner12}.  
The presence of multiple shock velocities in this region is confirmed by the very detailed studies of H$_2$ emission summarized in \citet{goldsmith11}, 
Different shock--related molecules observed in \peakone\ do not have their emission maxima at the same position \citep{goicoechea14}, so that we should not necessarily expect to find an exact coincidence between the maximum of the \otwo\ emission and that of other species. 

\section{Summary}
We have presented {\it Herschel} observations of \otwo\ toward Orion \peaka, a small source 6\as\ west of Orion IRc2.  The line features of the \otwo\ transitions at 487~GHz and 774~GHz are similar to the results from the \peakone\ observations with a \vlsr\ of 10--11~\kms\ and a line width of 2--3~\kms, but the 1121 GHz line is not convincingly detected toward \peaka. CH$_3$OCHO is identified to be an interloper largely or completely responsible for the feature that would otherwise be \otwo\ at a velocity 5--6~\kms\ in the 487~GHz data.

Among all the molecular lines identified in the {\it Herschel} spectra, SO is the only species showing a similar 11~\kms\ narrow component toward \peakone. Since SO is a sensitive tracer of shocks, the 11 \kms\ component could be tracing the same region in which the abundance of \otwo\ has been enhanced.  NO, which has a chemistry similar to that of \otwo, was found in previous work to have a narrow 11 \kms\ velocity component as well.  The intensity ratios of \otwo\ lines, especially of the 774 and 1121 GHz lines towards the \peaka\ and \peakone\ positions, suggest that the emitting source is not \peaka\ and likely to be close to \peakone, which is a region heated by recent passage of a shock.  The association of the \otwo\ emission with the shock is supported by detection of 621 GHz water maser features in this region, which have a velocity of $\simeq$ 12 \kms, very close to that of the \otwo.  

The best--fit LTE models indicate a small source size $\simeq$ 9\as\ FWHM, a low kinetic temperature $\sim30$ K, and a peak \otwo\ column density $\simeq$ 1.1$\times$10$^{18}$ cm$^{-2}$.  We have run models of a 12 \kms\ shock propagating into a condensation assumed to have had preshock density 2$\times$10$^4$ \cmv.  This condensation is modeled as having passed through a preshock evolution in which atomic and molecular species have, to different degrees, been depleted onto dust grain surfaces, but we find that the preshock history of the region only modestly  affects the postshock chemical evolution.  What is critical is having a reasonable UV radiation field ($\chi$ $\geq$ 1) present during the passage of the shock.  An \otwo\ fractional abundance exceeding 10$^{-5}$ with maximum value a factor 2--5 times greater (and possibly approaching 10$^{-4}$ if the sputtering efficiency is higher than we have assumed), can be produced for a period of 1--3~$\times$10$^4$ yr.  

Even with this high abundance and postshock density of 2~$\times$10$^4$ \cmv, the required line--of--sight dimension of the \otwo--enhanced region is significantly greater than its plane--of--the--sky dimension, but this geometry may be reflected in the 12~\kms\ velocity of water masers seen in this region having their maximum gain along the line of sight.  The relatively short period during which the \otwo\ fractional abundance is enhanced means that, while shocks are known to exist within many star-forming molecular clouds, the fraction of sources with significantly enhanced $X$(\otwo) will be relatively small. This is consistent with the generally low \otwo\ abundances that have been found in star--forming molecular clouds.

Toward the Hot Core, we derive a 1$\sigma$ upper limit to \xotwo\ of 0.9-1.5$\times$10$^{-7}$ assuming a temperature between 150 K and 300 K.  
A region in which dust and gas are at temperatures above 100~K (as can likely be produced by heating from nearby IR sources in the vicinity) can eventually develop a large gas--phase abundance of \otwo, as long as the UV flux is low.
However, establishment of this asymptotic result of pure gas--phase chemistry can take in excess of 10$^5$ yr.  This may well be longer than the lifetime of such regions, explaining why we do not see a dramatically enhanced \otwo\ abundance in the Hot Core, and why the \otwo\ abundance is generally very low in regions of massive star formation.

\section{Appendix - Modeling of Source Coupling}
\label{Appendix}

To more quantitatively constrain the location and size of the \otwo--emitting region, we assume that it is a single source having a Gaussian 1/$e$ width $\theta_{\rm s}$.
The brightness distribution is then
\begin{equation}
I_{\nu}(\Omega) = I_{\nu}(0)exp[-(\theta/\theta_{\rm s})^2] \lc
\end{equation}
where $\theta$ is the angle from the center of the emitting source.
If we assume that the source has a peak column density of \otwo\ molecules in the upper level of the transition being observed equal to $N_u$ and that the emission is optically thin, the peak brightness of the source is
\begin{equation}
I_{\nu} (0) = \frac{A_{\rm ul}hcN_{\rm u}}{4\pi\delta V} \lp
\end{equation}

In general, the antenna temperature resulting from observation of the source with an antenna having effective area $A_{\rm e}$ is given by
\begin{equation}
\label{ta_general}
T_{\rm A} = \frac{A_{\rm e}}{2k} \int\int P_{\rm n}(\Omega)I_{\nu}(\Omega) d\Omega \lc
\end{equation}
where $P_{\rm n}$ is the normalized antenna power pattern.  
We assume that the main beam of the antenna pattern is a Gaussian having 1/$e$ width $\theta_{\rm b}$, so that for the main beam, $P_{\rm n}~=~exp[-(\theta/\theta_{\rm b})^2]$.

In the limit of a source uniformly filling the entire antenna pattern, the source brightness can be assumed constant and taken out of the integral, which then becomes $\int$$\int$$P_{\rm n} d\Omega$ over all solid angle.  
The result is the antenna solid angle $\Omega_{\rm A}$. 
From the antenna theorem, $A_{\rm e} \Omega_{\rm A}$~=~$\lambda ^2$, which yields in this limit
\begin{equation}
T_{\rm A} = \frac{\lambda^2}{2k}I_{\nu}(0) \lp
\end{equation}
This ``ideal'' relationship defines an antenna temperature, which is useful in calculating what is observed in more realistic situations.
The main beam solid angle is defined as $\Omega_{\rm mb}$~=~$\int$$\int_{\rm mb}P_{\rm n}d\Omega$, and for the assumed Gaussian form is equal to $\pi\theta_{\rm b}^2$.
The main beam efficiency is $\epsilon_{\rm mb}~=~\Omega_{\rm mb}/\Omega_{\rm A}$.  
If we assume that the source is large compared  to the main beam size ($\theta_{\rm s}$ $\gg$ $\theta_{\rm b}$), but does not couple significantly to the rest of the antenna pattern (which is plausible given that much of the power pattern outside the main beam is due to diffraction from the secondary and the support legs, and hence is located at much larger angles than $\theta_{\rm b}$), the power received and the antenna temperature are multiplied by the factor $\Omega_{\rm mb}/\Omega_{\rm A}$, so that
\begin{equation}
T_{\rm A}(main~beam~filled) = \frac{\Omega_{mb}}{\Omega_{\rm A}} \frac{\lambda^2}{2k}I_{\nu}(0) = \epsilon_{\rm mb} T_{\rm A} = \frac{\Omega_{\rm mb}}{\Omega_{\rm A}} T_{\rm A}\lp
\end{equation}

In the limit of small source size ($\theta_{\rm s}$ $\ll$ $\theta_{\rm b}$),  we can consider that the normalized response is equal to unity over the source, and can be taken out of the integral, which is then equal to the source solid angle, $\Omega_{\rm s}~=~\pi\theta_{\rm s}^2$.
The antenna temperature is then
\begin{equation}
T_{\rm A}(small~source) = \frac{A_{\rm e} \Omega_{\rm s}}{2k} I_{\nu}  \lp
\end{equation}
Again using the antenna theorem, we obtain
\begin{equation}
T_{\rm A}(small~source) = \frac{\Omega_{\rm s}}{\Omega_{\rm A}}\frac{\lambda^2}{2k}I_{\nu}(0) = \frac{\Omega_{\rm s}}{\Omega_{\rm A}} T_{\rm A}\lp
\end{equation}

Returning to the general case given by Equation \ref{ta_general}, we see that the antenna temperature is proportional to the convolution of two Gaussians (again considering only the main beam of the antenna pattern), which itself is a Gaussian.
We define $\theta_{\rm al}$ by
\begin{equation}
\frac{1}{\theta_{\rm al}^2} = \frac{1}{\theta_{\rm s}^2} + \frac{1}{\theta_{\rm b}^2} \lc
\end{equation}
which defines the aligned source solid angle
\begin{equation}
\Omega_{\rm al} = \pi \theta_{\rm al}^2 \lp
\end{equation}
For the case of the beam direction offset from that of the source by angle $\theta_0$, the angle $\theta_{\rm co}$ is the 1/$e$ width of the two convolved Gaussians
\begin{equation}
\label{theta_co}
\theta_{\rm co}^2 = \theta_{\rm s}^2 + \theta_{\rm b}^2 \lp
\end{equation}
Equation \ref{ta_general} then yields
\begin{equation}
\label{Ta_1}
T_{\rm A}(obs) = T_{\rm A} \frac{\Omega_{\rm al}}{\Omega_{\rm A}} exp[-(\theta_0/\theta_{\rm co})^2] \lp
\end{equation}

With the assumption that there is a single source characterized by a particular value of $I_{\nu}(0)$ and $\theta_{\rm s}$, then when observed with a given beam size $\theta_{\rm b}$ it would, if the telescope direction were offset by angle $\theta_0$ from the source direction, produce a particular value of $T_A(obs)$.
The two observations have offset angles $\theta_{\rm 0A}$ and $\theta_{01}$ from the \peaka\ and \peakone\ positions, respectively.  
The ratio of the antenna temperatures for the two observations is then
\begin{equation}
\label{ratio1}
\frac{T_{\rm A}(obs~A)}{T_{\rm A}(obs~1)} = \frac{exp[-(\theta_{\rm 0A}/\theta_{\rm co})^2]}{exp[-(\theta_{01}/\theta_{\rm co})^2]} \lc
\end{equation}
since the coupling factor due purely to the beam and source sizes cancels out.
The ratio in equation \ref{ratio1} applies equally well to the integrated intensity; the preceding formulas are also applicable to the integrated as well as the peak antenna temperature.

\begin{figure}
\begin{center}
\includegraphics[width=0.9\textwidth,angle=270]{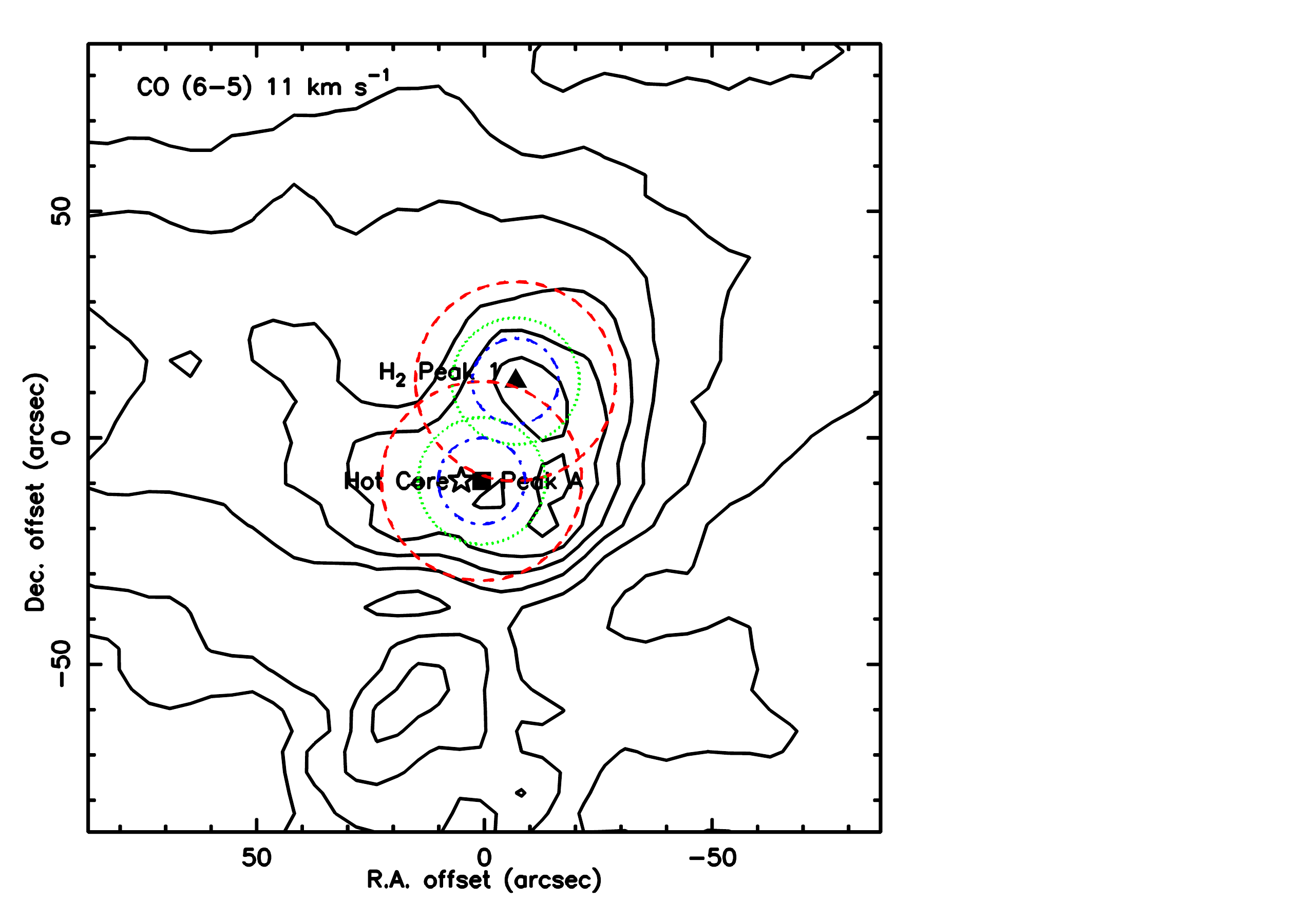}
\caption{Contours of CO(6-5) emission in the 11 {\kms} channel of width 1 $\kms$ centered on the Orion BN source ($\alpha_{2000} = 5^h35^m14^s.16, \delta_{2000} = -5\degree22\am21\as.5$) from \citet{peng12}. The position of \peaka\ ($\alpha_{2000} = 5^h35^m14^s.2$, $\delta_{2000} = -5\degree22\am31\as$) is indicated by the black square. \peakone\ ($\alpha_{2000} = 5^h35^m13^s.7, \delta_{2000} = -5\degree22\am09\as$) is indicated by the black triangle. The star indicates the position of the Hot Core, and the Compact Ridge is located $\sim$10\as\ southwest of the Hot Core. The circles indicate the FWHM beam sizes at the three observed frequencies: 44\as\ at 487 GHz (red dashed lines), 28\as\ at 774 GHz (green dotted lines), and 19\as\ at 1121 GHz (blue dot-dashed lines). }
\label{fig:co}
\end{center}
\end{figure}

\begin{figure}
\begin{center}
\includegraphics[width=0.5\textwidth]{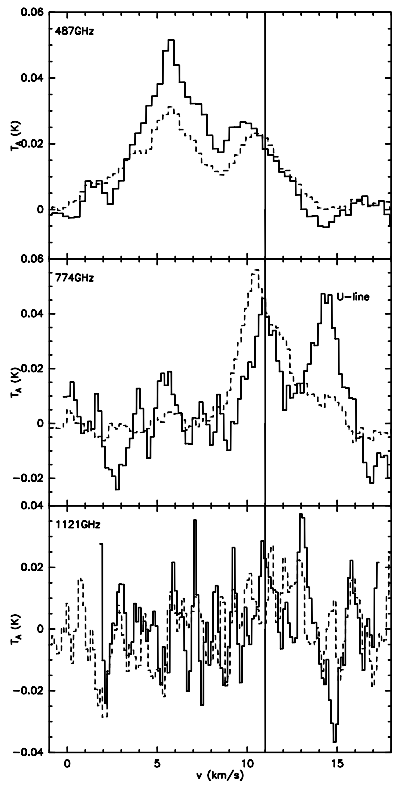}
\caption{Spectra of the three rotational transitions of {\otwo} toward \peaka\ are shown by the solid lines. The dashed lines are the same spectra toward \peakone\ position from \citet{goldsmith11}.}
\label{fig:o2}
\end{center}
\end{figure}

\begin{figure}
\begin{center}
\includegraphics[width=0.8\textwidth]{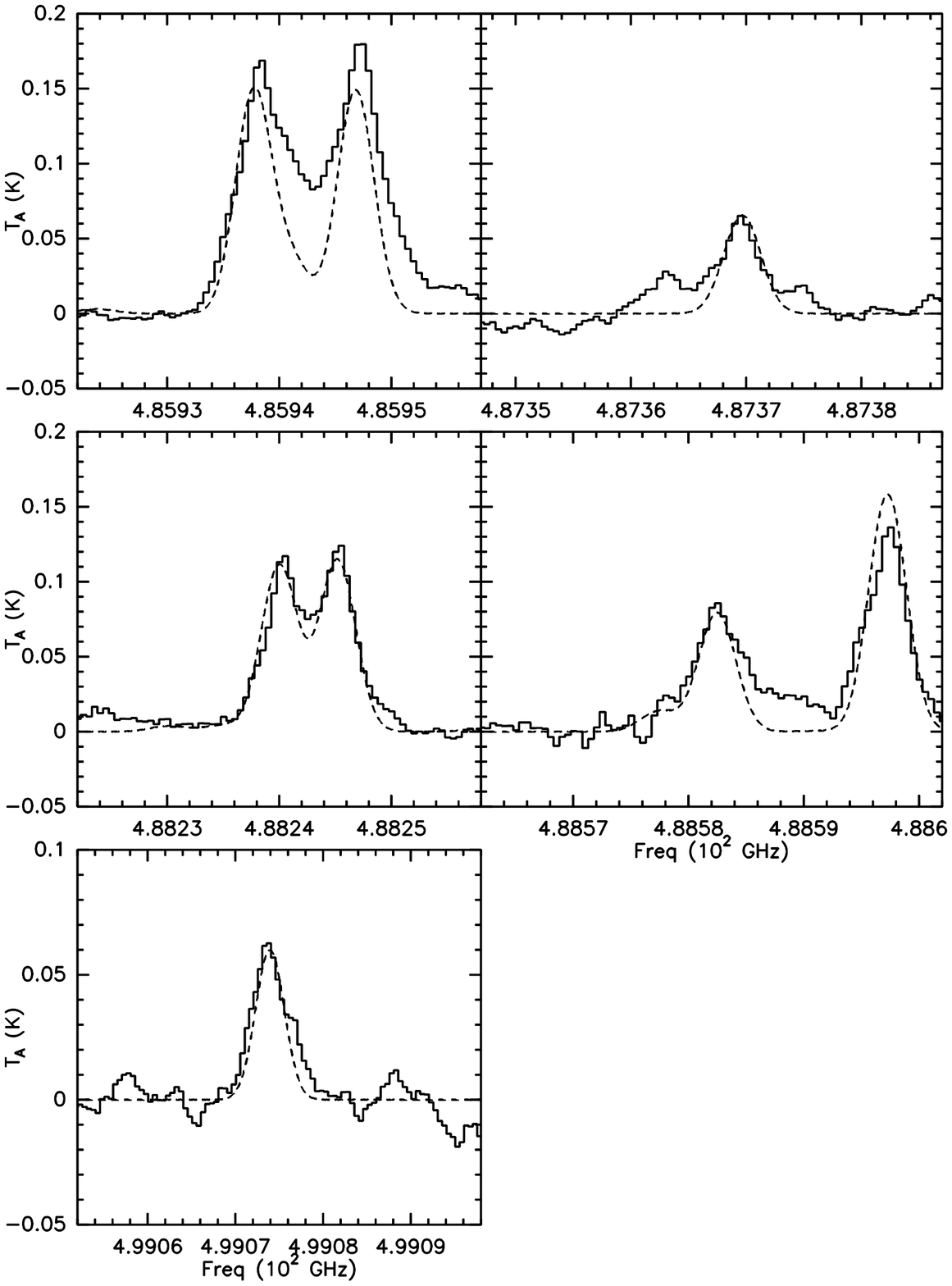}
\caption{Modeled CH$_3$OCHO lines (dashed curves) having upper level energies similar to that of  the \otwo\ 487 GHz transition plotted together with the observed lines toward \peaka\ (solid curves).}
\label{fig:mf_peakA}
\end{center}
\end{figure}

\begin{figure}
\begin{center}
\includegraphics[width=0.8\textwidth]{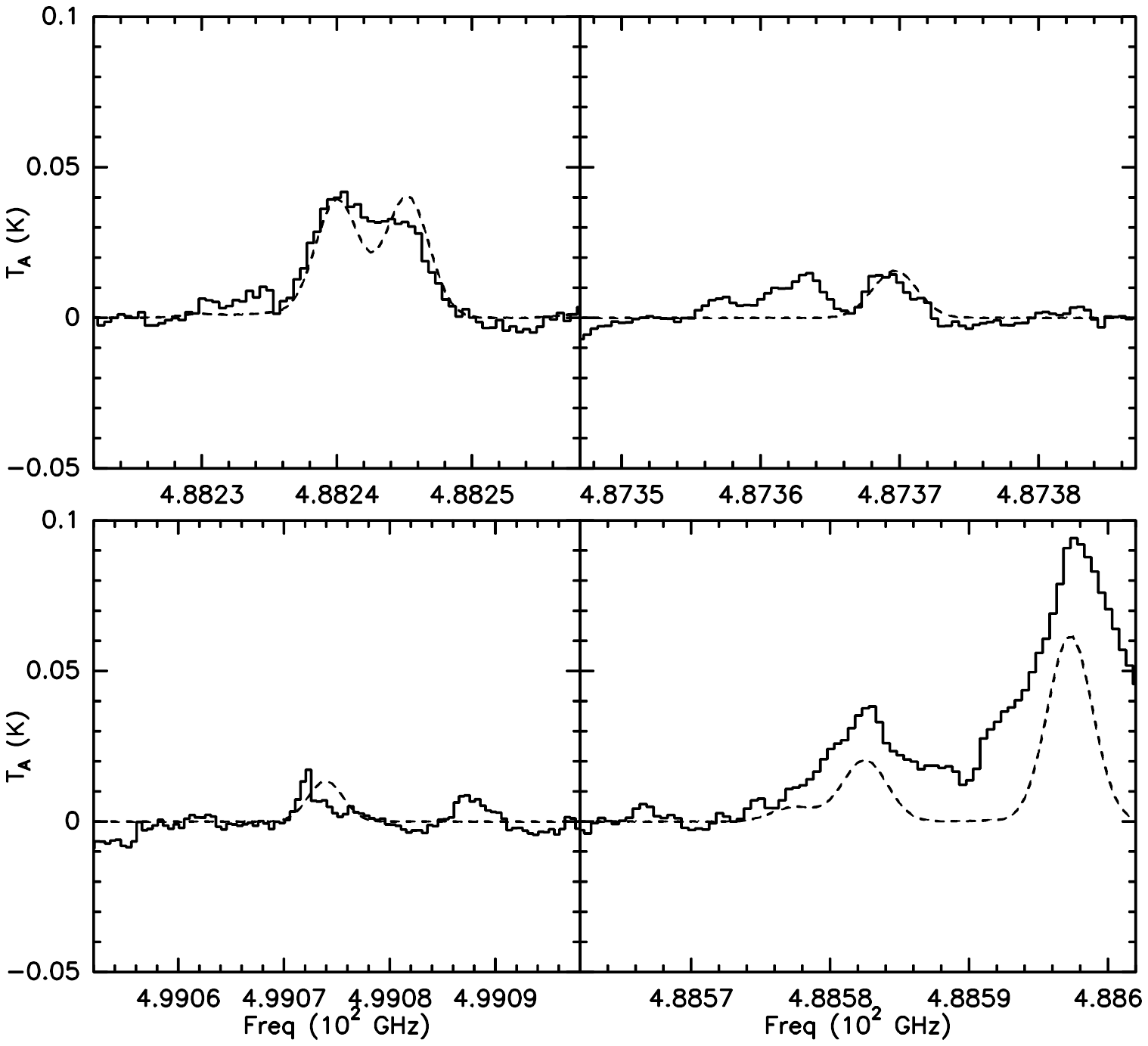}
\caption{Modeled CH$_3$OCHO lines (dashed curves) having similar upper level energies simialr to that of the \otwo\ 487 GHz transition plotted with the observed lines toward \peakone\ (solid curves).}
\label{fig:mf_peak1}
\end{center}
\end{figure}

\begin{figure}
\begin{center}
\includegraphics[width=0.5\textwidth]{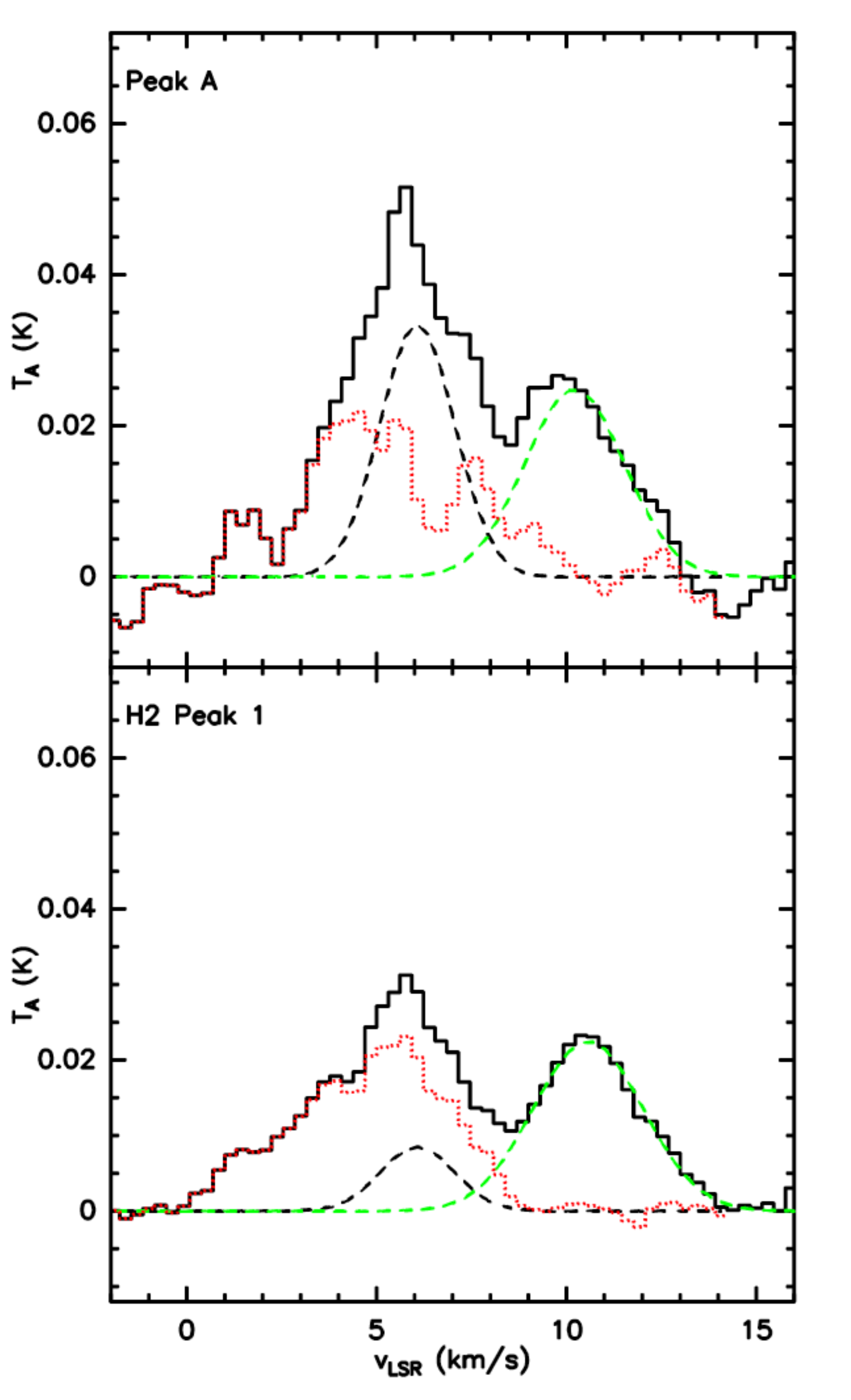}
\caption{Observed spectra (solid curves) together with modeled CH$_3$OCHO line (dashed curve) towards \peaka\ (upper panel) and \peakone\ (lower panel). The velocity scale is for the \otwo\ 487 GHz transition.  The green dashed lines are the Gaussian fits to the 11 \kms\ \otwo\ line. The residual (\otwo\ and CH$_3$OCHO lines subtracted) is indicated by red dotted lines.  The intensity of the CH$_3$OCHO line at the two positions is derived from the fits of the other lines observed simultaneously as shown in Figures \ref{fig:mf_peakA} and \ref{fig:mf_peak1}.}
\label{fig:residual}
\end{center}
\end{figure}

\begin{figure}
\begin{center}
\includegraphics[width=0.8\textwidth]{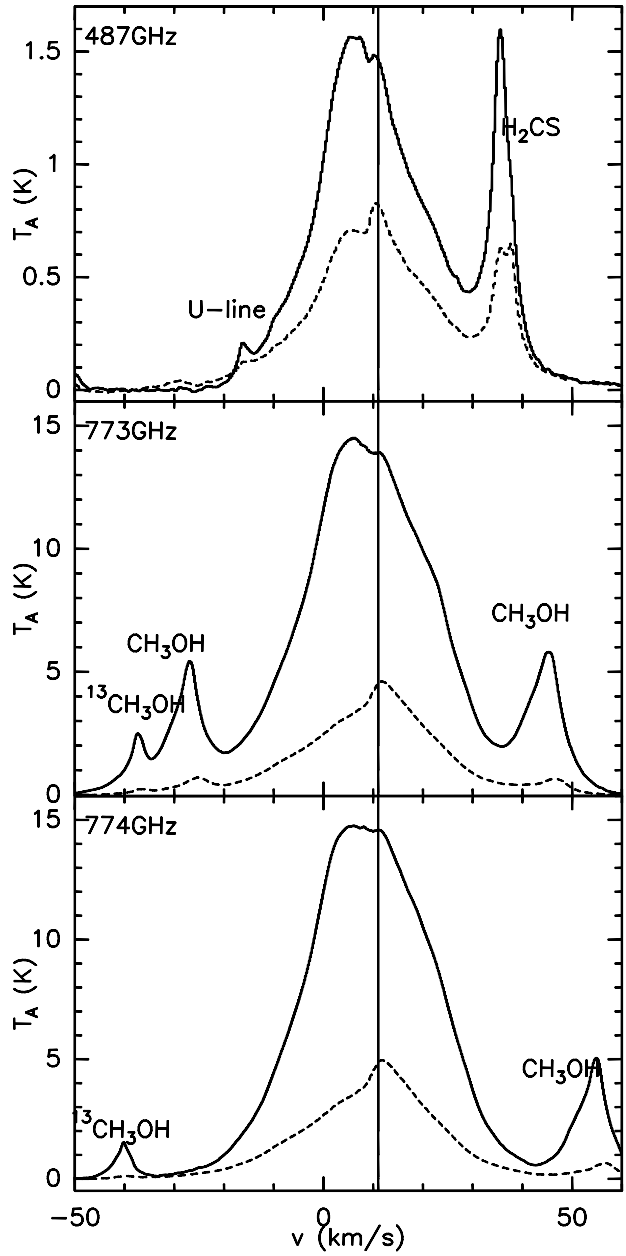}
\caption{SO 7$_7$--7$_6$ transition at 487.71 GHz (upper panel), 17$_{18}$--16$_{17}$ transition at 773.5 GHz (middle panel), and 19$_{18}$--18$_{17}$ transition at 774.45 GHz (lower panel) toward \peaka\ (solid curves) and \peakone\ (dashed curves).  The vertical line indicates a velocity of 11 \kms\ for SO.}
\label{fig:so}
\end{center}
\end{figure}

\begin{figure}
\begin{center}
\includegraphics[width=0.9\textwidth]{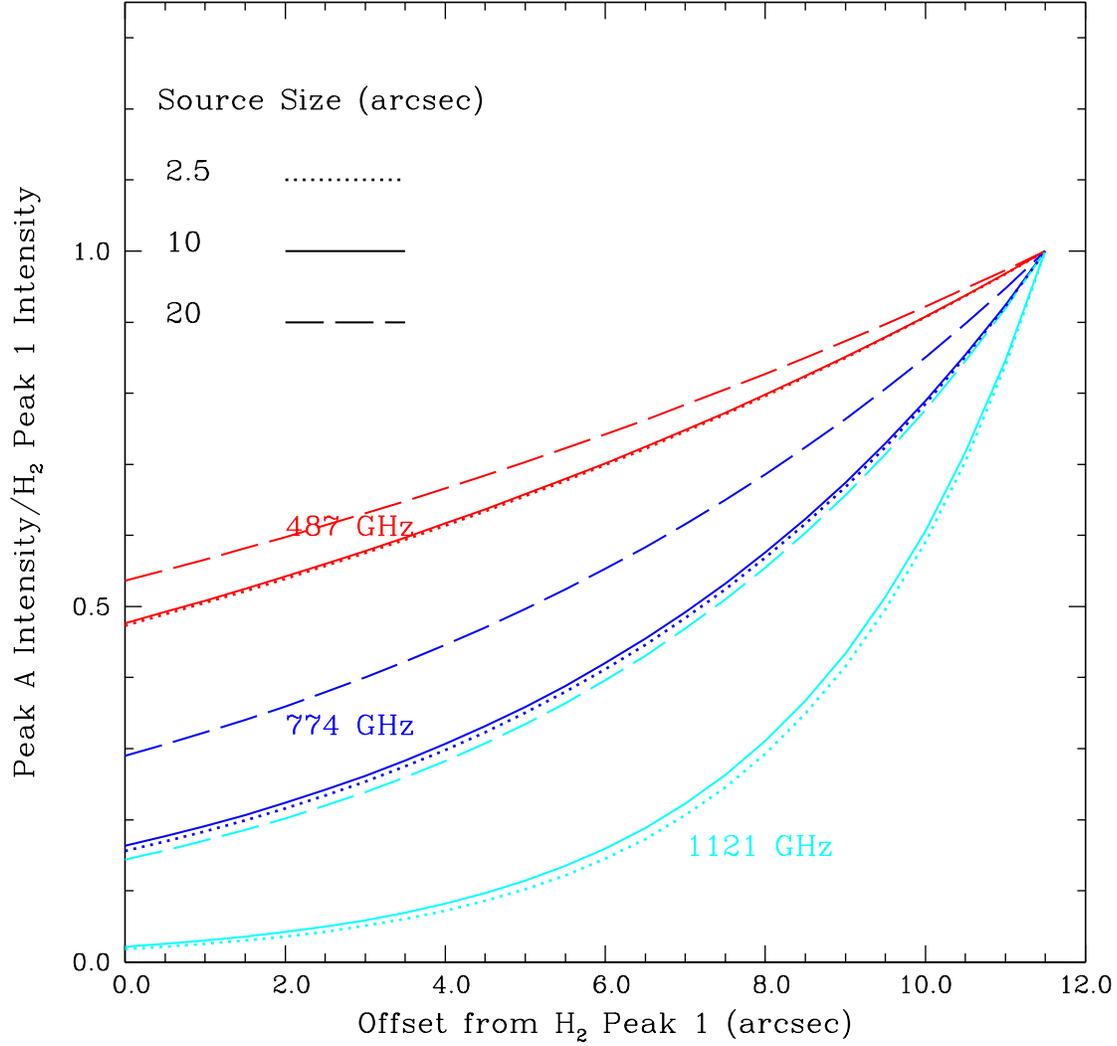}
\caption{\label{fig:pkApk1_rel} Integrated intensities of the three \otwo\ lines observed towards  \peaka\ (new data) compared to those previously observed pointed at \peakone.  The three curves are for the indicated values of the FWHM source size, that correspond to $\theta_c$ = 1.5\as, 6.0\as, and 12.0\as, respectively.  The horizontal axis is the offset of the source position relative to that of \peakone. \peaka\ is 23\as\ from the position of \peakone, so that if the source is offset by 11.5\as\ from \peakone, the observed intensity ratios will be unity.}
\end{center}
\end{figure}

\begin{figure}
\begin{center}
\includegraphics[width=0.8\textwidth, angle = 270]{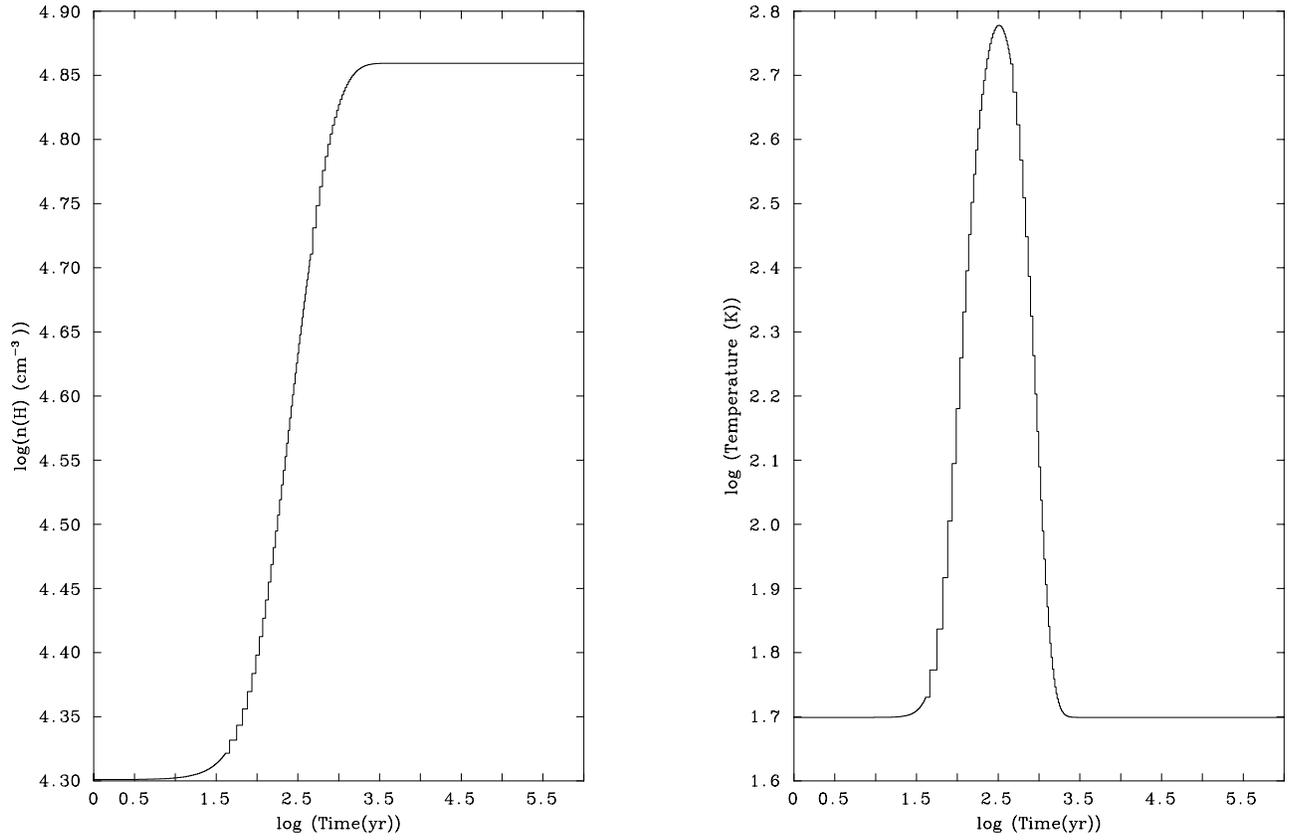}
\caption{\label{fig:dens-temp}  Profiles of density (left) and  temperature (right) produced by the 12 \kms\ shock propagating into a region with preshock H$_2$ density equal to 2$\times$10$^4$ \cmv.  For this run, the postshock minimum temperature (as well as the initial Phase II temperature) is constrained to be above a minimum value of 50 K, while the maximum temperature of the shocked gas is 600 K.}
\end{center}
\end{figure}

\begin{figure}
\begin{center}
\includegraphics[width=0.8\textwidth, angle = 270]{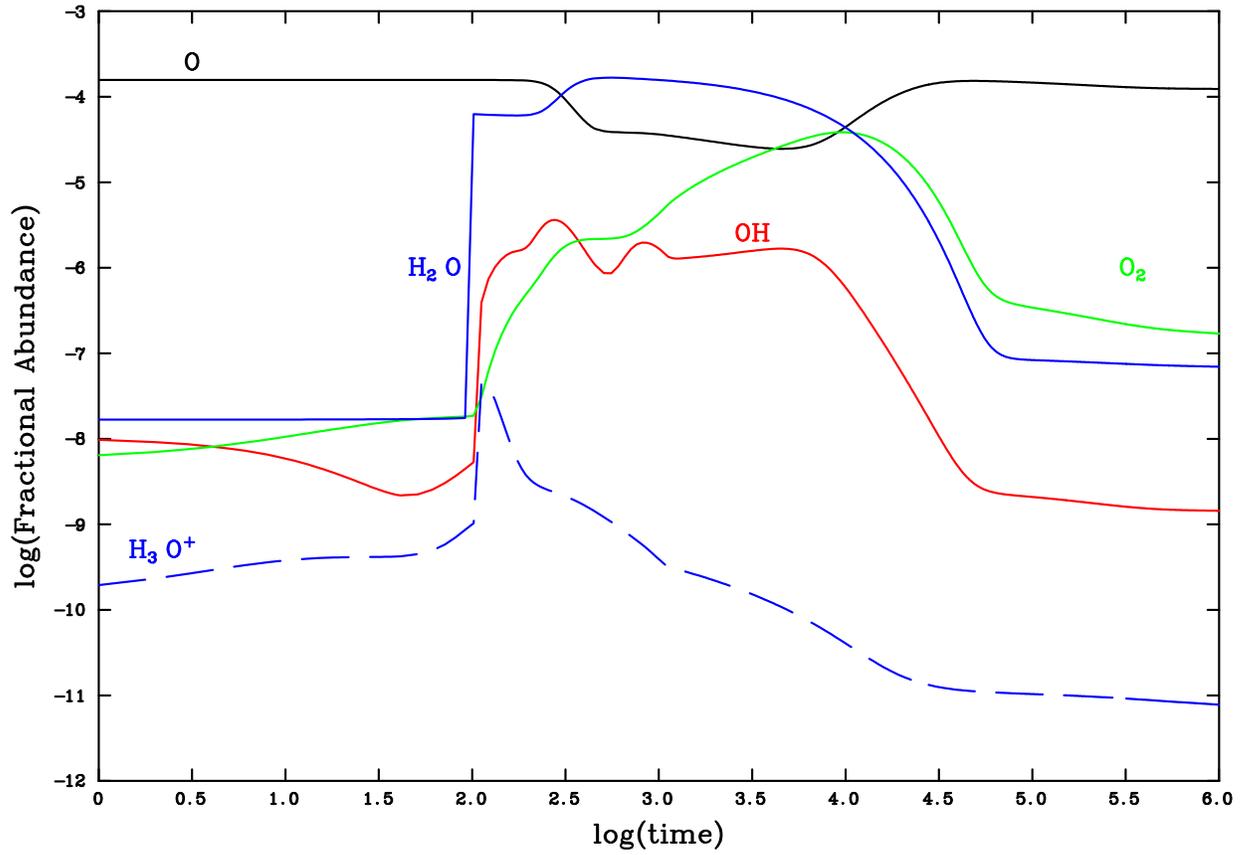}
\caption{\label{1uvshort}  Abundance of key species as function of time in Phase II, for external UV field with $\chi$ = 1 (flux equal to standard Draine value) and ``short'' Phase I with relatively little grain surface depletion before the shock. The horizontal axis is the logarithm of the time (in yr) after passage of the shock front.  X(O$_2$) above 10$^{-5}$ is maintained for 3$\times$10$^4$ years following the passage of the shock front.}
\end{center}
\end{figure}

\begin{figure}
\begin{center}
\includegraphics[width=0.8\textwidth, angle = 270]{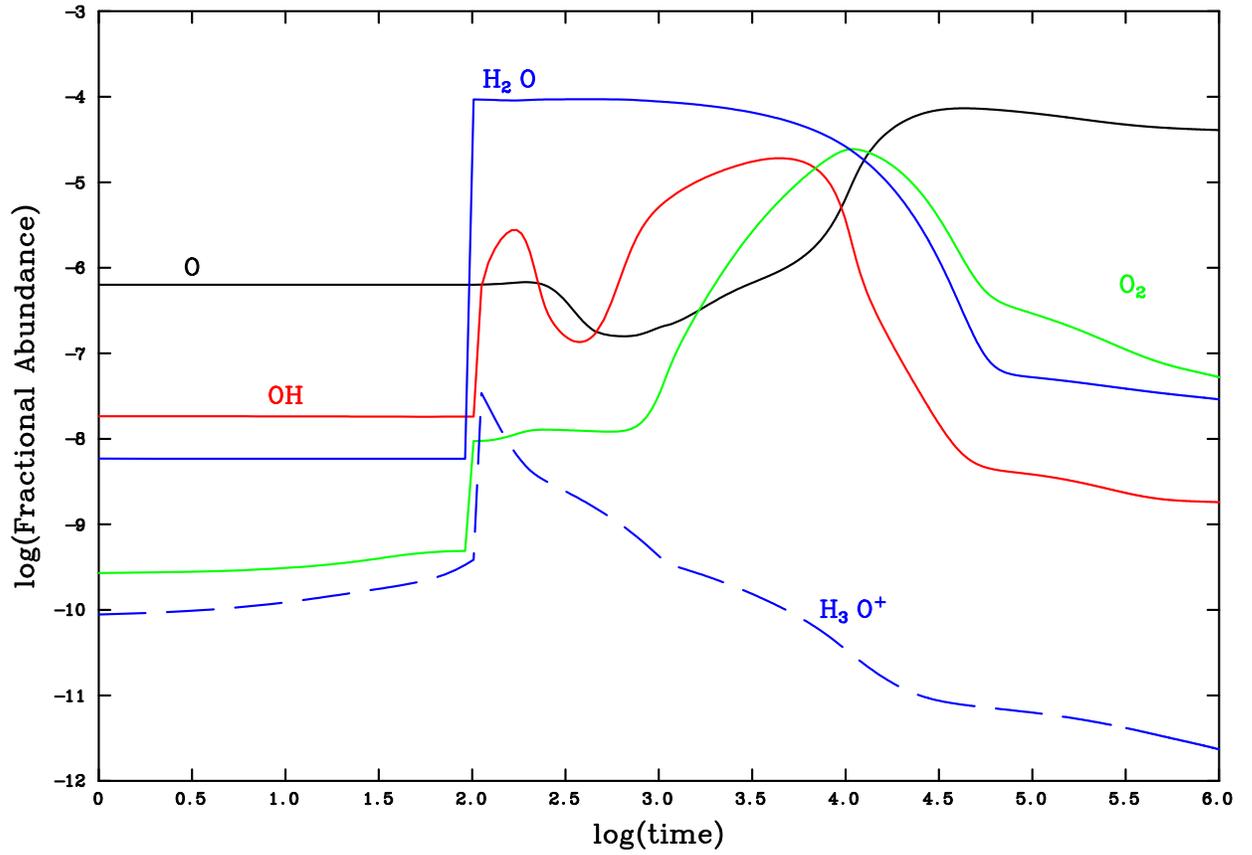}
\caption{\label{1uvlong}  As Figure \ref{1uvshort} but with  ``long'' Phase I with relatively more grain surface depletion before the shock. The horizontal axis is the logarithm of the time (in yr) after passage of the shock front. The general behavior of the O$_2$ abundance is seen to be independent of the preshock conditions determined by the duration of Phase I. While the O$_2$ fractional abundance takes longer after the shock to rise,  X(O$_2$) between 10$^{-5}$ and 10$^{-4}$ is again maintained for up to 3$\times$10$^4$ years following the passage of the shock front.}
\end{center}
\end{figure}

\begin{figure}
\begin{center}
\includegraphics[width=0.8\textwidth, angle = 270]{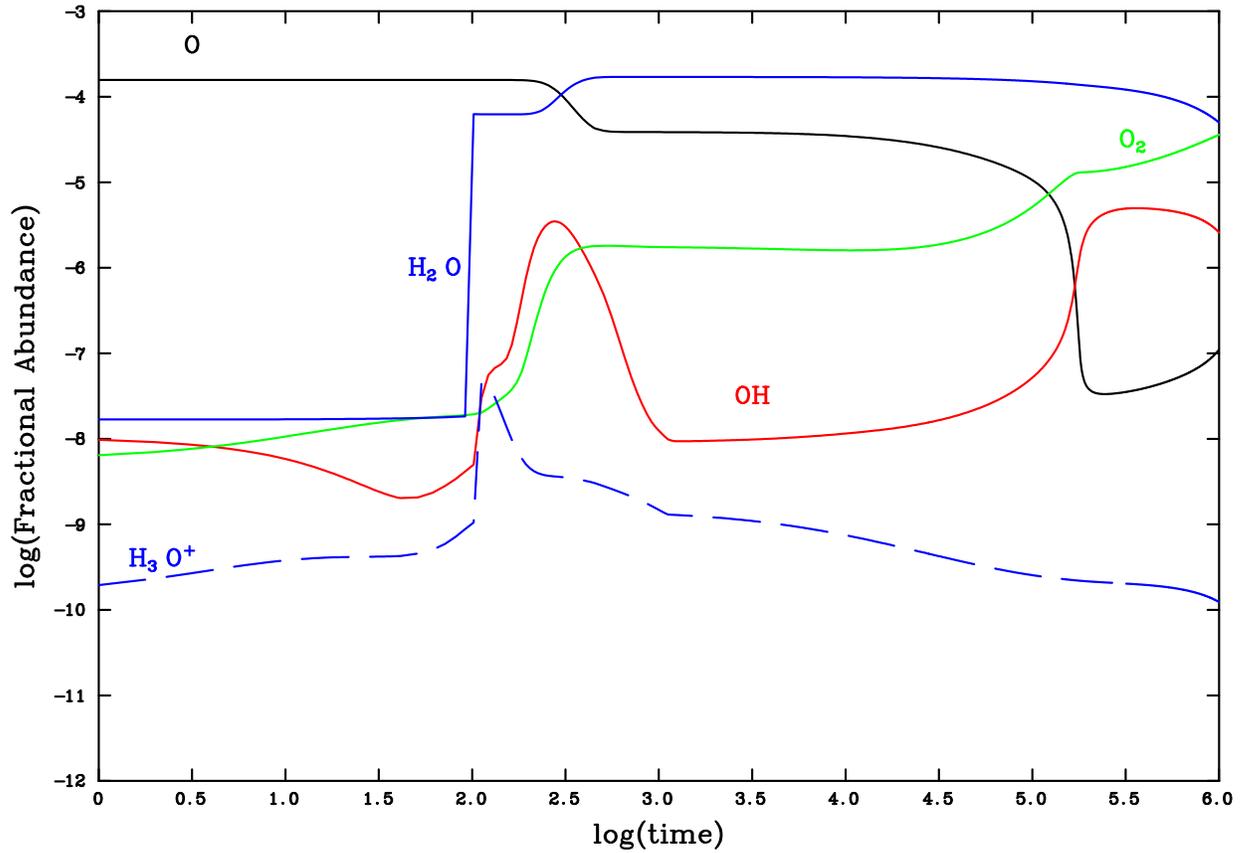}
\caption{\label{0uv}  Abundances of key species as a function of time in Phase II, with no external UV field  and a ``short'' Phase I.    The horizontal axis is the logarithm of the time (in yr) after passage of the shock front. X(\otwo) exceeds 10$^{-6}$ only for times more than 3$\times$10$^4$ yr following passage of the shock front.  The \otwo\ fractional abundance is still rising slowly 10$^6$ yr after passage of the shock front as OH and \otwo\ become more abundant than atomic oxygen, and the abundance of gas--phase water begins to drop.}
\end{center}
\end{figure}

\begin{figure}
\begin{center}
\includegraphics[width = 0.8\textwidth,angle = 270]{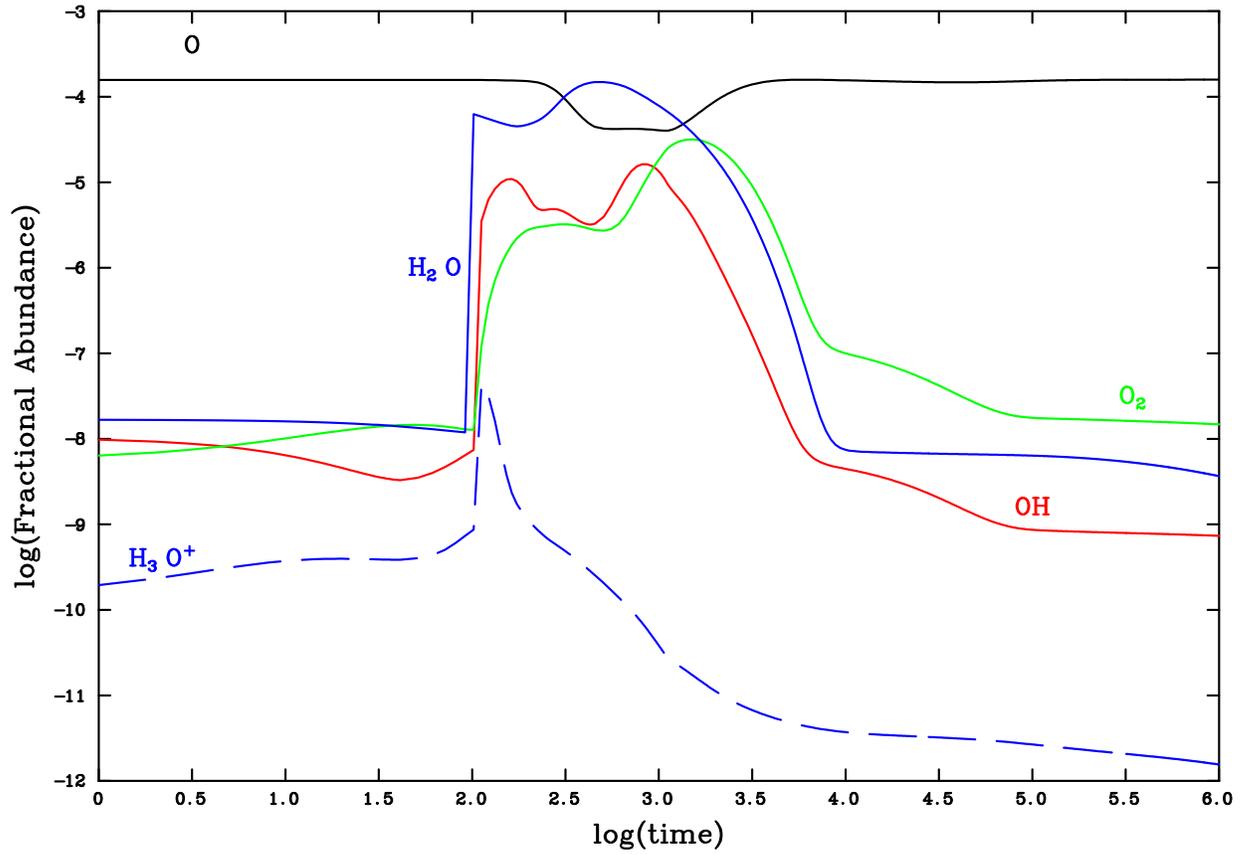}
\caption{\label{10uv}  Abundances of key species as a function of time in Phase II, for external UV field with $\chi$ = 10 for a ``short'' Phase I.    The horizontal axis is the logarithm of the time (in yr) after passage of the shock front. X(O$_2$) greater than 10$^{-5}$  is present in the interval 0.8 to 3$\times$10$^4$ yr following the passage of the shock front.}
\end{center}
\end{figure}

\begin{figure}
\begin{center}
\includegraphics[width = 0.8\textwidth, angle = 270]{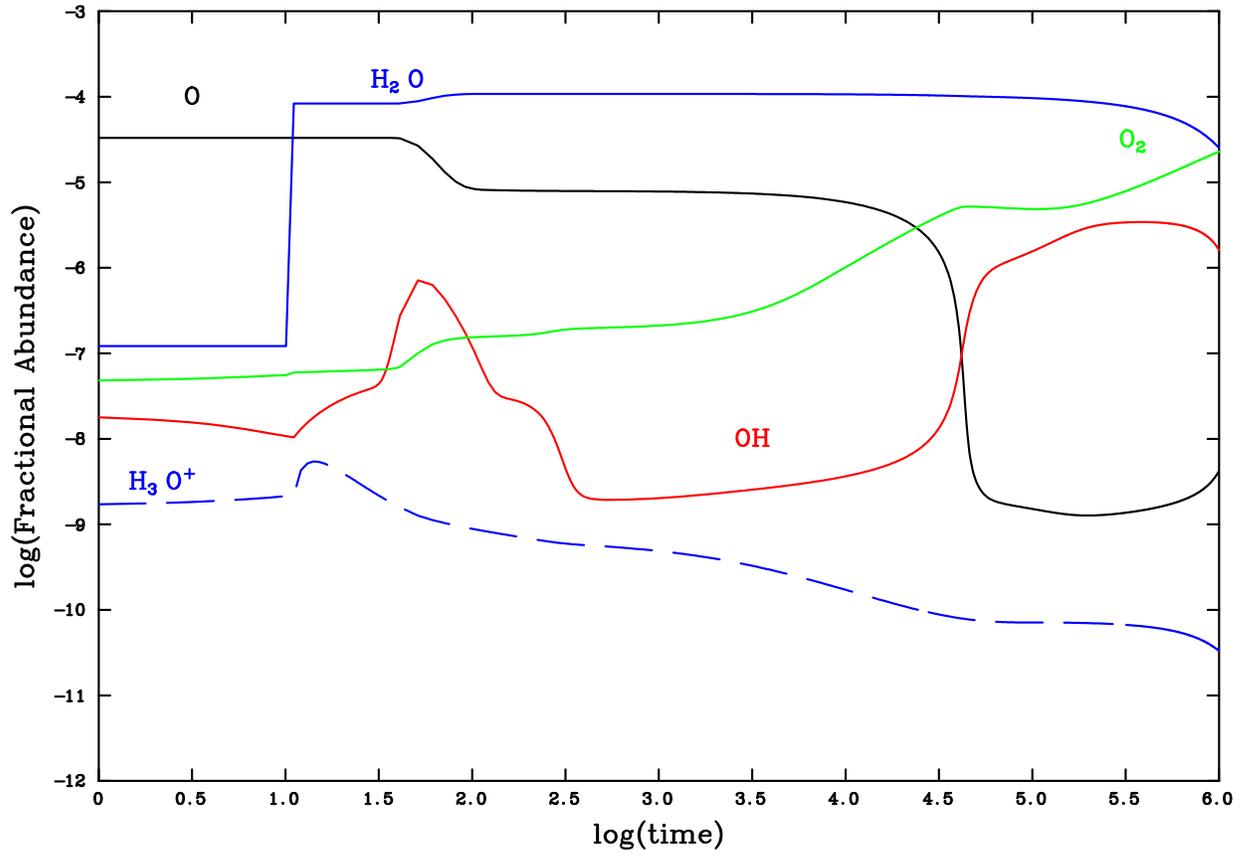}
\caption{\label{1uv10^5}  Abundances of key species as a function of time in Phase II, for all parameters as in Figure~\ref{1uvshort} except that the density of the preshock gas is 10$^5$ cm$^{-3}$.   The horizontal axis is the logarithm of the time (in yr) after passage of the shock front. The main effect is that the factor of 10 increase in the extinction results in dramatic reduction of photo rates and thus largely mimics a situation in which there is no external UV radiation field.}
\end{center}
\end{figure}

\acknowledgements
HIFI has been designed and built by a consortium of institutes and university departments from across Europe, Canada and the United States (NASA) under the leadership of SRON, Netherlands Institute for Space Research, Groningen, The Netherlands, and with major contributions from Germany, France and the US. This work was carried out in part at the Jet Propulsion Laboratory, California Institute of Technology.  JRG thanks the Spanish MINECO for funding support under grants CSD2009-00038, AYA2009-07304, and AYA2012-32032.  We thank Shiya Wang, Nathan Crockett, and the NASA Herschel Science Center Helpdesk for the help with the data reduction. We appreciate the help from Justin Neill with installing and using XCLASS. We thank Tzu-Cheng Peng for providing the CO map used in Figure \ref{fig:co}. We thank the anonymous reviewer for a number of suggestions that improved the clarity of the paper.


\newpage
\clearpage
\begin{deluxetable}{rcrllrr}
\tablecolumns{7}
\tabletypesize{\tiny}
\tablecaption{Observed \otwo\ lines \label{table:observations}}
\tablewidth{0pt}
\tablehead{
\colhead{Frequency\tablenotemark{a}} & \colhead{Transition\tablenotemark{a}} & \colhead{\Eu\tablenotemark{a}} & Date & OBSIDs & H-pol offset & V-pol offset\\
\colhead{(MHz)} & \colhead{} & \colhead{(K)} & & & (arcsec) & (arcsec)
}
\startdata
487249.38 	& $N$=3--1, $J$=3--2 & 26 & OD 1065 & 1342244299, 1342244300, 1342244302, 1342244303 & (+1.8, +2.7) & ($-$1.8, $-$2.8)\\
			&				   &	   &		     &	1342244301, 1342244305,  1342244306 & (+1.8, +2.8) & ($-$1.8, $-$2.8)\\ 
			& 				   & 	   &                & 1342244304						& (+1.8, +2.8) & ($-$1.8, $-$2.7)\\
773839.69 	& $N$=5--3, $J$=4--4 & 61 & OD 1065 & 1342244289, 1342244290, 1342244292, 1342244294, 1342244295 & ($-$0.1, +2.3) & (+0.2, $-$2.3)\\ 
			& 				   & 	   &		     & 1342244291, 1342244293, 1342244296 & ($-$0.2, +2.3) & (+0.1, $-$2.3)\\ 
1120715.04	& $N$=7--5, $J$=6--6 & 115 & OD 1203 & 1342250406, 1342250407, 1342250408 & ($-$1.4, $-$0.3) & (+1.3, +0.3)\\
			&				   &	   & OD 1205 & 1342250447, 1342250448 & ($-$1.4, $-$0.2) &  (+1.4, +0.3)\\
			&				&	&		& 1342250449 & ($-$1.3, $-$0.2) & (+1.4, +0.3)\\
&&& OD 1206 & 1342250459 & ($-$1.3, $-$0.2) & (+1.4, +0.3)\\
&&& OD 1224 & 1342251197 & ($-$1.4, $-$0.0) & (+1.4, +0.0)\\
\enddata
\tablenotetext{a}{JPL Line Catalog \citep{pickett98}}
\end{deluxetable}

\begin{deluxetable}{rlrcr}
\tablecolumns{5}
\footnotesize
\tablecaption{Parameters of the observed \otwo\ lines at the \peaka\ position \label{table:parameters}}
\tablewidth{0pt}
\tablehead{
\colhead{Freq} & \colhead{I($\sigma$)\tablenotemark{a}} & \colhead{ \vlsr\ ($\sigma$)} & \colhead{$\Delta$v($\sigma$)} & \colhead{T$_{A}$}\\
 \colhead{GHz} & \colhead{(K~\kms)} & \colhead{(\kms)} & \colhead{(\kms)} & \colhead{(K)}
}
\startdata
487 & 0.081(0.011) & 10.22(0.19) & 3.08(0.39) & 0.025\\
       & 0.178(0.012) & 5.75(0.11) & 3.77(0.31) & 0.044\\
774 & 0.074(0.009) & 10.97(0.10) & 1.74(0.25) & 0.040\\
1121 & 0.022(0.007) & 11.02(0.14) & 0.82(0.32) & 0.024\\
\enddata
\tablenotetext{a}{statistical errors}
\end{deluxetable}


\begin{deluxetable}{ccccccccccc}
\tablecolumns{11}
\tabletypesize{\tiny}
\footnotesize
\tablecaption{Uncertainties in the \otwo\ integrated intensities (K\,km\,s$^{-1}$) \label{table:uncertainties}}
\tablewidth{0pt}
\tablehead{
\colhead{} & \colhead{} & \colhead{} & \colhead{H$_2$ Peak 1} & \colhead{} & \colhead{} & \colhead{} & \colhead{} & \colhead{Peak A} & \colhead{} & \colhead{}\\
\colhead{Freq} &\colhead{I} & \colhead{Gaussian Fit} & \colhead{Baseline} & \colhead{Calibration\tablenotemark{a}} & \colhead{Combined} & \colhead{I} & \colhead{Gaussian Fit} & \colhead{Baseline} & \colhead{Calibration\tablenotemark{a}} & \colhead{Combined}\\
 \colhead{(GHz)} &\colhead{} & \colhead{Uncert.} & \colhead{Uncert.} & \colhead{Uncert.} & \colhead{Uncert.} & \colhead{} & \colhead{Uncert.} & \colhead{Uncert.} & \colhead{Uncert.} & \colhead{Uncert.}\\
}
\startdata
487   &  0.081 & 0.004 & 0.016 & 0.008 & 0.018 & 0.081 & 0.011 & 0.016 & 0.008 & 0.021\\
774   &  0.154 & 0.004 &    -      & 0.015 & 0.016 & 0.074 & 0.009 & 0.015 & 0.007 & 0.019\\
1121 &  0.043 & 0.009 &    -      & 0.004 & 0.010 & 0.022 & 0.007 & 0.020 & 0.002 & 0.021\\
\enddata
\tablenotetext{a}{We assume a 10\% calibration uncertainty.}
\end{deluxetable}
\clearpage

\begin{deluxetable}{cccrr}
\tablecolumns{5}
\tabletypesize{\tiny}
\tablecaption{Modeled CH$_3$OCHO lines\tablenotemark{a} \label{table:mf}}
\tablewidth{0pt}
\tablehead{
\colhead{Species} & \colhead{Frequency} & \colhead{Transition} & \colhead{Log$_{10}A_{ul}$} & \colhead{\Eu}\\
\colhead{} & \colhead{(MHz)} & \colhead{J'$_{K'_a,K'_c}$--J$_{K_a,K_c}$} & \colhead{} & \colhead{(K)}
}
\startdata
CH$_3$OCHO v=1 & 487252.00 & $40_{ 6,34}-39_{ 6,33}$ E & -2.35 & 704.7\\
\hline
CH$_3$OCHO v=1 & 485923.80 & $41_{ 6,36}-40_{ 5,35}$ A & -3.40 & 714.9\\
CH$_3$OCHO v=0 & 485937.73 & $39_{ 8,32}-38_{ 8,31}$ E & -2.34 & 508.8\\
CH$_3$OCHO v=1 & 485941.17 & $39_{ 9,31}-38_{ 9,30}$ A & -2.34 & 703.5\\
CH$_3$OCHO v=0 & 485946.82 & $39_{ 8,32}-38_{ 8,31}$ A & -2.34 & 508.8\\
CH$_3$OCHO v=1 & 485948.87 & $22_{ 6,16}-21_{ 4,17}$ A & -4.84 & 361.3\\
\hline
CH$_3$OCHO v=1 & 487369.67 & $44_{ 2,42}-43_{ 3,41}$ A & -2.98 & 737.1\\
CH$_3$OCHO v=1 & 487369.67 & $44_{ 3,42}-43_{ 3,41}$ A & -2.46 & 737.1\\
CH$_3$OCHO v=1 & 487369.67 & $44_{ 2,42}-43_{ 2,41}$ A & -2.46 & 737.1\\
CH$_3$OCHO v=1 & 487369.68 & $44_{ 3,42}-43_{ 2,41}$ A & -2.98 & 737.1\\
\hline
CH$_3$OCHO v=1 & 488230.08 & $20_{18, 2}-20_{17, 3}$ A & -4.35 & 527.2\\
CH$_3$OCHO v=1 & 488230.08 & $20_{18, 3}-20_{17, 4}$ A & -4.35 & 527.2\\
CH$_3$OCHO v=1 & 488234.18 & $18_{18, 0}-18_{17, 1}$ A & -4.76 & 504.4\\
CH$_3$OCHO v=1 & 488234.18 & $18_{18, 1}-18_{17, 2}$ A & -4.76 & 504.4\\
CH$_3$OCHO v=1 & 488235.20 & $19_{18, 1}-19_{17, 2}$ A & -4.49 & 515.5\\
CH$_3$OCHO v=1 & 488235.20 & $19_{18, 2}-19_{17, 3}$ A & -4.49 & 515.5\\
CH$_3$OCHO v=0 & 488240.03 & $39_{11,29}-38_{11,28}$ E & -2.35 & 544.0\\
CH$_3$OCHO v=1 & 488243.64 & $27_{18, 9}-27_{17,10}$ E & -2.96 & 626.0\\
CH$_3$OCHO v=1 & 488245.19 & $30_{18,13}-30_{17,14}$ E & -3.21 & 676.8\\
CH$_3$OCHO v=0 & 488245.23 & $39_{11,29}-38_{11,28}$ A & -2.35 & 544.0\\
CH$_3$OCHO v=1 & 488246.64 & $40_{22,18}-39_{22,17}$ A & -2.49 & 991.6\\
CH$_3$OCHO v=1 & 488247.02 & $40_{22,19}-39_{22,18}$ A & -2.49 & 991.6\\
\hline
CH$_3$OCHO v=0 & 488577.43 & $41_{ 6,36}-40_{ 5,35}$ E & -3.38 & 531.0\\
CH$_3$OCHO v=1 & 488582.44 & $45_{ 1,44}-44_{ 2,43}$ E & -4.90 & 741.6\\
CH$_3$OCHO v=1 & 488582.46 & $45_{ 1,44}-44_{ 1,43}$ E & -2.35 & 741.6\\
CH$_3$OCHO v=1 & 488582.46 & $45_{ 2,44}-44_{ 2,43}$ E & -2.35 & 741.6\\
CH$_3$OCHO v=1 & 488582.49 & $45_{ 2,44}-44_{ 1,43}$ E & -4.90 & 741.6\\
CH$_3$OCHO v=0 & 488582.68 & $41_{ 6,36}-40_{ 5,35}$ A & -3.38 & 531.0\\
CH$_3$OCHO v=1 & 488591.18 & $18_{ 2,16}-17_{ 1,17}$ A & -5.67 & 292.7\\
CH$_3$OCHO v=0 & 488597.28 & $38_{ 9,29}-37_{ 9,28}$ E & -2.32 & 498.0\\
\hline
CH$_3$OCHO v=1 & 499073.93 & $46_{ 1,45}-45_{ 2,44}$ E & -4.85 & 765.5\\
CH$_3$OCHO v=1 & 499073.95 & $46_{ 1,45}-45_{ 1,44}$ E & -2.33 & 765.5\\
CH$_3$OCHO v=1 & 499073.95 & $46_{ 2,45}-45_{ 2,44}$ E & -2.33 & 765.5\\
CH$_3$OCHO v=1 & 499073.98 & $46_{ 2,45}-45_{ 1,44}$ E & -4.85 & 765.5\\
\enddata
\tablenotetext{a}{From JPL Line Catalog \citep{pickett98}}
\end{deluxetable}

\begin{deluxetable}{rl}
\tablecolumns{2}
\footnotesize
\tablecaption{Ratios of \otwo\ line intensities at the two observed positions 
\label{table:pkApk1_relative}}
\tablewidth{0pt}
\tablehead{
\colhead{Freq} & \colhead{$I_{\rm Peak\,A}$/$I_{\rm H_2\,Peak\,1}$\tablenotemark{a}}\\
 \colhead{(GHz)} & \colhead{}
}
\startdata   
487 & 1.00$\pm$0.31\\
774 & 0.48$\pm$0.13\\
1121 & 0.51$\pm$0.51\\
\hline
\enddata
\tablenotetext{a}{Combined statistical errors, baseline uncertainties, and calibration uncertainties.}
\end{deluxetable}

\begin{deluxetable}{ccc}
\tablecolumns{3}
\tablewidth{0pt}
\tablecaption{\label{table:ratios} Ratios of \otwo\ integrated intensities at the two observed positions\tablenotemark{a} \label{table:ratio}}
\tablehead{\colhead{Ratio} & \colhead{Peak A}  & \colhead{H$_2$ Peak 1}}
\startdata
487/774  & 1.10$\pm$0.37  & 0.53$\pm$0.13\\
1121/774 & 0.30$\pm$0.29  & 0.28$\pm$0.07\\
\enddata
\tablenotetext{a} {Not corrected for coupling of antenna beam to the source}
\end{deluxetable}
\clearpage

\begin{deluxetable}{cccc}
\tablecolumns{4}
\footnotesize
\tablecaption{Peak \otwo\ column density from models with different source sizes \label{table:densities}}
\tablewidth{0pt}
\tablehead{
\colhead{Source Size\tablenotemark{a}} & \colhead{$\chi^2$~\tablenotemark{b}} & \colhead{Kinetic Temperature} & \colhead{Peak \otwo\ Column Density\tablenotemark{c}}\\
\colhead{(arcsec)} & \colhead{} & \colhead{(K)} & \colhead{(cm$^{-2}$)}
}
\startdata
20		& 0.90 & 32 & $2.9\times10^{17}$\\
10		& 0.36 &32 & $8.8\times10^{17}$\\
9		& 0.14 &31 & $11.1\times10^{17}$\\
8 	&0.21 &31 &$13.8\times10^{17}$\\
5  & 0.65 &31 &$3.4\times10^{18}$\\
2.5&1.41 & 30 &$1.62\times10^{19}$\\
\enddata
\tablenotetext{a}{The FWHM source size}
\tablenotetext{b}{Defined as $\sum$[(observed$-$model line intensities)$^2$/(the observed uncertainty)$^2$]}
\tablenotetext{c}{The peak \otwo\ column density in a pencil beam}
\end{deluxetable}

\begin{deluxetable}{ccc}
\tablecolumns{3}
\footnotesize
\tablewidth{0pt}
\tablecaption{\label{table:reactions} Dominant pathways for formation and destruction of H$_2$O, O, OH, and O$_2$ in the model illustrated in Figure~\ref{1uvshort}, for different time periods after passage of the shock front}
\tablehead{\colhead{Reaction} & \colhead{Percentage} & \colhead{Time after Passage of Shock Front (yr)}}
\startdata
 & & \\
\multicolumn{3}{c}{H$_2$O} \\
\hline
H + OH $\rightarrow$ H$_2$O + $h\nu$ & 40-50 & 5  \\
\hline 
H$_3$O$^+$ +   HCN $\rightarrow$ H$_2$O + HCNH$^+$ & 30-50 & $<$ 90  \\
\hline
NH$_3$ + H$_3$O$^+$ $\rightarrow$ H$_2$O + NH$_4^+$  & 70 &  100-120 \\ 
\hline
H$_2$ + OH $\rightarrow$ H$_2$O + H &  80-100 & 130-800 \\
\hline
NH$_2$ + NO $\rightarrow$ H$_2$O  +  N$_2$  & 20-70 & $>$ 800 \\
\hline
H$_2$O + $h\nu$ $\rightarrow$ OH + H & 60 & $\leq$ 100 \\
\hline
C$^+$ + H$_2$O $\rightarrow$  HCO$^+$ + H & 40 &  $\leq$ 100 \\
\hline
H$_2$O + $h\nu$ $\rightarrow$ OH + H & 100 & $\geq$ 100 \\
\hline
& & \\
\multicolumn{3}{c}{O} \\
\hline
CO + $h\nu$ $\rightarrow$ O + C & 50-90 & $<$ 100 \\
\hline
OH + $h\nu$ $\rightarrow$ O + H & 40-60 & 100--150 \\
\hline
O$_2$ + $h\nu$ $\rightarrow$ O + O & 40-90 & 1300--10$^5$ \\
\hline
O + OH $\rightarrow$ O$_2$ + H & 50--90 & $\leq$ 10 and $>$ 750, $<$ 3$\times$10$^4$ \\ 
\hline
O + H$_2$ $\rightarrow$ O + OH &  20-50 & 180--200 \\  & 80-90 & 200--500 \\
\hline
& & \\
\multicolumn{3}{c}{OH} \\
\hline
O + H$_2$CO $\rightarrow$ OH + CO &45-55 & $\geq$ 100 \\
\hline
H$_2$O + $h\nu$ $\rightarrow$ OH + H & 60-100&  100--200 and 600--4$\times$10$^4$ \\
\hline
H$_2$ + O $\rightarrow$ OH + H & 60-90 & 200--420 \\
\hline
O + OH $\rightarrow$ O$_2$ + H & 60-97 & $\geq$ 100 and $>$ 600 \\
\hline
H$_2$ + OH $\rightarrow$ H$_2$O + H & 60-97 & 100-600 \\
\hline
& & \\
\multicolumn{3}{c}{O$_2$} \\
\hline
O + OH $\rightarrow$  O$_2$ + H & 100 & always \\
\hline
O$_2$ + $h\nu$ $\rightarrow$ O + O & 40-90 & $\leq$ 2$\times$10$^5$ \\
\hline
O$_2$ + S $\rightarrow$ SO + O & 40-60 & $>$ 8$\times$10$^4$ \\
\hline
\enddata
\end{deluxetable}
\clearpage


\clearpage

\clearpage

\clearpage

\clearpage
%
%
\clearpage
%
%
\clearpage
%
%
\clearpage

\clearpage

\clearpage

\clearpage

\clearpage

\clearpage

\clearpage

\bibliography{140530}
\end{document}